\documentclass[11pt]{article}
\usepackage{caption}
\usepackage{lmodern, anyfontsize}
\usepackage[T1]{fontenc}
\usepackage[utf8]{inputenc}
\usepackage{amsmath, amssymb, bm, mathrsfs, mathtools, hyperref, lipsum}
\usepackage[numbered]{bookmark}
\usepackage{booktabs}
\usepackage{mdframed, comment}
\usepackage{courier, lineno}

\usepackage{authblk}
\hypersetup{
    colorlinks,
    citecolor=black,
    filecolor=black,
    linkcolor=black,
    urlcolor=black
}
\usepackage[letterpaper, margin=0.70in]{geometry}
\usepackage{graphicx, float, multicol, balance}
\usepackage[version=4]{mhchem}
\usepackage[sorting=none, maxnames=50, style=science]{biblatex}
\title{Detrital Remanent Magnetization Analysis at the Particle Scale: Implications for Prebiotic Origins of Homochirality}

\author{
S. Furkan Ozturk$^{1}$\thanks{Correspondence should be addressed to S.F.O. at ozturk@caltech.edu}, 
Roger R. Fu$^{2}$,
José A. P. M. Devienne$^{3}$,
Sammy Piascik$^{2}$, 
Perri Thaler$^{2}$,
Aaron T. Kuan$^{3}$,
Richard J. Harrison$^{4}$,
Dimitar D. Sasselov$^{5}$\\

\footnotesize{$^{1}$Division of Geological and Planetary Sciences, California Institute of Technology, Pasadena, CA 91125, USA}\\

\footnotesize{$^{2}$Department of Earth and Planetary Sciences, Harvard University, Cambridge, MA 02138, USA}\\

\footnotesize{$^{3}$Yale School of Medicine, Yale University, New Haven, CT 06510, USA}\\

\footnotesize{$^{4}$Department of Earth Sciences, University of Cambridge, Cambridge CB2 3EQ, UK}\\

\footnotesize{$^{5}$Department of Astronomy, Harvard University, Cambridge, MA 02138, USA}}

\date{\today}

\begin{document}
\maketitle

\begin{abstract}
Detrital remanent magnetization (DRM), the alignment of ferromagnetic grains with the geomagnetic field during sedimentation, is the primary mechanism of paleomagnetic remanence acquisition in sedimentary rocks, yet its microphysical mechanism remains contested. Measured DRM intensities consistently fall below theoretical predictions---the long-standing ``DRM efficiency gap''---and no single-particle resolving observations have been available to test DRM acquisition models directly. Here we apply quantum diamond microscopy (QDM) to laboratory-redeposited lacustrine sediments to obtain the first single-particle magnetizations of DRM carriers. We compare these measurements with the results of bulk magnetometry and a Langevin-based kinetic model. We observe a statistically significant mean alignment of individual detrital particle moments with the applied field, 10\% when averaged as unit vectors and 24\% when weighted by moment magnitude, with the vector sum of all particles reproducing the direction of the bulk DRM, which is $\sim$1.8\% of saturation remanence. Particle-level analysis shows that alignment rises with particle moment, a trend reproduced by our kinetic model in which floc-hosted magnetite rotates toward the field until the surrounding sediment locks it in place. Even where post-depositional disturbance is minimal and the measured DRM reaches the upper end of reported efficiencies, it remains well below its theoretical maximum, limited in comparable measure by incomplete particle-level alignment and sub-saturation within-particle magnetization. Beyond its implications for the DRM mechanism, this strong particle-level bias provides a quantitative foundation for magnetic minerals as natural chiral agents, a proposed pathway for the emergence of biomolecular homochirality on the prebiotic Earth.\\
\end{abstract}


\section{Introduction}

Detrital remanent magnetization (DRM) is acquired as ferromagnetic grains, free or hosted within silicate-dominated particle agglomerates, settle through a water column under the influence of the ambient geomagnetic field and come to rest at the sediment--water interface (\autoref{fig:fig1}a)~\cite{dunlop2001rock, TauxeBook}. It is the dominant mechanism of remanence acquisition in fine-grained clastic sediments and the physical basis for nearly all sedimentary paleomagnetic records of plate motions and relative paleointensity variations~\cite{TauxePaleointensties}. The classical theoretical framework predicts that a dilute suspension of non-interacting single-domain particles in thermal equilibrium with the geomagnetic field acquires a remanence following the Langevin function of the dimensionless magnetic-to-thermal energy ratio, with alignment efficiencies of tens of percent for single-domain ferrimagnetic carriers in Earth-strength fields \cite{nagata1961rock, Tauxe1993}.

Experimentally, however, redeposition studies consistently find DRM intensities orders of magnitude below this equilibrium prediction and roughly linear in applied-field strength over the geomagnetic range~\cite{TauxeSteindorf2006, MitraTauxe2009, RobertsHuang2020}. The microphysical origin of this ``DRM efficiency gap'' remains contested, with two broad mechanistic classes dominating the literature. In the first, flocculation binds ferromagnetic carriers to hydrodynamically much larger silicate or organic aggregates whose rotational dynamics are governed by inertial and mechanical rather than magnetic torques, so that individual carriers never reach their equilibrium alignment during settling~\cite{TauxeSteindorf2006, ShcherbakovSycheva2010, Heslop2007}. In the second, syn- and post-depositional processes---particle--substrate collision, rolling on the sediment surface, bioturbation, and compaction-driven axial rotation during burial---shallow the recorded inclination and reduce the net moment relative to the suspended state~\cite{King1955, TauxeKent1984, Bilardello2013, PierceSwansonHysell2022, TauxeBook}. Because conventional magnetometers integrate over $10^6$--$10^{10}$ carriers of unknown size, shape, and mineralogy, discriminating between these mechanisms has been difficult, and the microphysics of DRM acquisition remains a central open question in sedimentary paleomagnetism~\cite{TauxePaleointensties}.

The strength and spatial extent of natural remanent magnetization in natural sediments, mainly detrital and chemical remanent magnetization, is also relevant to the prebiotic origin of biological homochirality. Homochirality---the single-handedness of the chiral molecules of life, such as $L$-amino acids and $D$-sugars---is one of the enduring open questions of the origin of life, and its origins on the early Earth requires both a strong chiral symmetry-breaking agent and a mechanism to amplify an induced imbalance to homochirality~\cite{blackmond2019origin}. Many such agents have been proposed, including circularly polarized ultraviolet light, parity-violating weak interactions, and enantiomorphic mineral surfaces; each, however, provides either an exceedingly small enantiomeric bias, operates only locally, or lacks a plausible mechanism for propagating the bias across the full prebiotic network~\cite{sallembien2022possible, ozturk2023central, Ozturk2025Homochirality}. Magnetized magnetite surfaces have been shown to act as chiral agents under ambient conditions through the chiral-induced spin selectivity (CISS) effect \cite{bloom2024chiral}, templating enantioselective crystallization of ribose aminooxazoline (RAO), a central RNA precursor, to homochirality \cite{ozturk2023central, ozturk2023crystallization}, which can then lead to a homochiral network~\cite{ozturk2023central, Ozturk2025Homochirality}. Recent work has highlighted natural remanence mechanisms, in particular chemical remanent magnetization (CRM), as an efficient way for prebiotic magnetite grains to maintain the chirality-induced magnetization in lacustrine environments \cite{Devienne2026}. For this magnetization-based mechanism to have initiated the emergence of homochirality, however, a sufficient proportion of magnetized particles at the water-sediment interface must have been aligned to the geomagnetic field so that they broke the chiral symmetry across the length and time scales of the early-Earth geochemical environments that hosted prebiotic processes. Although the exact required threshold of particle-level alignment is still being studied, very low alignment efficiencies of 0.1 - 1$\%$ reported in natural DRM-carrying sediments likely pose a significant challenge to this model of homochirality originating from DRM \cite{HeslopRoberts2014}. 

Resolving these questions at the level of individual sedimenting particles has recently become possible through quantum diamond microscopy (QDM), a wide-field magnetic imaging technique that uses ensembles of nitrogen-vacancy centers in diamond as vector magnetometers with $\sim 1~\mu$m spatial resolution and $\sim 10^{-17}$~A\,m$^2$ moment sensitivity~\cite{GlennFu2017, fu2020high, QDMlab2022}. QDM has been applied to measure magnetic signals from individual chondrules to speleothems~\cite{GlennFu2017, piascik2025reliable}, has enabled paleointensity estimates from individual magnetic sources~\cite{bellon2026estimating}, and can discriminate secondary processes such as chemical remagnetization at the particle scale~\cite{qi2025quantifying}. No comparable analysis yet exists for sediments to understand how efficiently detrital carriers acquire remanence particle by particle. Applying QDM to soft sediments also requires embedding the carrier population in a matrix that preserves the depositional orientations at the micrometer scale, which has not previously been attempted.

Here we combine individual particle-level QDM with bulk magnetometry, first-order reversal curve (FORC) analysis, low-temperature magnetization measurements, and a Langevin-based kinetic model of DRM acquisition to resolve the microphysics of DRM in laboratory-redeposited lacustrine sediments fixed in resin. We discuss its implications for the long-standing DRM efficiency gap in paleomagnetism and for the prebiotic origin of biological homochirality.

\section{Results}

\subsection{Sediment collection and sample preparation}

Sediment was collected from two locations in Alewife Brook (Cambridge, MA, USA) and wet-sieved to remove large debris and coarser grains, isolating the fraction finer than 63~$\mu$m. The sieved material was washed in acetone and ethanol to dissolve and remove organic residues, and then redeposited in a controlled, sterile aqueous environment inside a low-magnetic-contamination glass jar. The jar was aligned in a known orientation relative to the local magnetic field in the lab ($|B| = 58.8~\mu\mathrm{T}$, inclination $+66.6^\circ$, declination $329.0^\circ$). 

\begin{figure}[H]
    \centering
    \includegraphics[width=\textwidth]{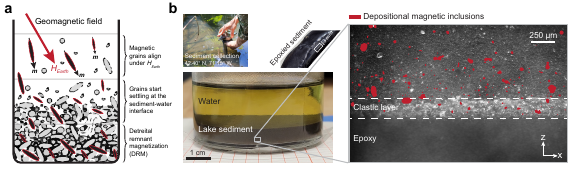}
    \caption{\textbf{Laboratory redeposition of lake sediments for single particle-level DRM analysis.}
    \textbf{a}, Schematic of depositional remanent magnetization (DRM) acquisition. Ferromagnetic grains suspended in a water column experience a magnetic torque from the local geomagnetic field $H$\textsubscript{Earth} that partially aligns their individual moments $\mathbf{m}$ with the field as they settle. The alignment is frozen in as the grains come to rest at the sediment--water interface, producing a net sedimentary remanence that records the field direction at the time of deposition.
    \textbf{b}, Lake sediment collected from Alewife Brook (Cambridge, MA, USA; 42.40$^\circ$N, 71.15$^\circ$W; top left) was sieved, redeposited in a glass jar oriented in the local geomagnetic field (center), and immobilized by graded solvent exchange into a two-part epoxy. A horizontal cross-section through the cured sample (right) exposes the water--sediment interface in the $x$--$z$ plane, with magnetic inclusions visible in red against the clastic layer and the underlying epoxy; scale bar 250~$\mu$m.}
    \label{fig:fig1}
\end{figure}

In order to immobilize magnetic particles we used a neural tissue imaging protocol for water-resin replacement \cite{kaub2026magnetic}. Once the suspension had settled under the ambient field, pore water was serially exchanged for ethanol, propylene oxide, and finally a two-part epoxy resin through a graded solvent series designed to minimize mechanical disturbance and to preserve the acquired DRM (\autoref{fig:fig1}b, Materials and Methods). The majority of the magnetic sources are concentrated in a thin depositional layer immediately above the coarser siliciclastic layer (\autoref{fig:fig1}b), consistent with settling-velocity separation during redeposition. Two independent redeposition experiments were performed. From the first, four samples were sectioned and polished for both bulk magnetometry and QDM analysis; from the second, one additional sample was analyzed by bulk magnetometry (\autoref{tab:drm_measurements}). All cutting, polishing, and mounting tools and adhesives used downstream of redeposition were verified to be free of magnetic contamination.

\subsection{Bulk magnetometry}

We first characterized the acquired DRM at the bulk sample scale using a 2G Enterprises Model 755 superconducting rock magnetometer (SRM), following the cutting and polishing protocol described in Materials and Methods. 

All five samples carried a well-resolved natural remanent magnetization (\autoref{tab:drm_measurements}, rows 2G1--2G5 and \autoref{fig:fig2}d). Across the four samples from the first redeposition experiment, declinations cluster within a few degrees of the local geomagnetic field, and inclinations are systematically shallower than the ambient $+66.6^\circ$ by $\sim25$--$30^\circ$ (flattening factor of $\sim0.35$)---the diagnostic signature of DRM inclination shallowing (\autoref{fig:fig2})~\cite{anson1987compaction, TauxeKent1984}. The sample from the second redeposition experiment (2G5) shows similar declination match but markedly less shallowing ($\sim 10^\circ$); in this experiment the epoxy solidified before a laminated post-depositional layer could form, producing a more well-mixed dispersion of dipoles in which the horizontal-bedding mechanisms responsible for shallowing were suppressed. Moment-amplitude variation across the table reflects differences in the analyzed sample volume. 

Finally, to quantify the bulk DRM efficiency, we applied a 1 T isothermal remanent magnetization (IRM) to the QDM-analyzed sample after all imaging was complete to retrieve a DRM to IRM ratio of 1:55 or 1.8$\%$, which is stronger than most literature reported DRM efficiencies \cite{Tauxe1993, roberts2013magnetic}. These bulk measurements confirm that both redeposition protocols record a consistent DRM, motivating the single-particle analysis that follows.

\begin{table}[ht]
\centering
\begin{tabular}{lrrrr}
\toprule
Measurement & Inclination ($^\circ$) & Declination ($^\circ$) & Moment amplitude (A\,m$^2$) & Shallowing ($^\circ$) \\
\midrule
2G1 & $+43.7$ & $322.5$ & $1.60 \times 10^{-9}$ & $22.9$ \\
2G2 & $+37.5$ & $332.8$ & $1.34 \times 10^{-8}$ & $29.1$ \\
2G3 & $+39.0$ & $334.5$ & $1.67 \times 10^{-9}$ & $27.6$ \\
2G4 & $+38.2$ & $325.7$ & $3.45 \times 10^{-9}$ & $28.4$ \\
2G5 & $+56.5$ & $326.4$ & $8.29 \times 10^{-9}$ & $10.1$ \\
\midrule
QDM1 & $+42.3$ & $329.2$ & $1.81 \times 10^{-12}$ & $24.3$ \\
QDM2 & $+26.3$ & $339.7$ & $1.49 \times 10^{-11}$ & $40.3$ \\
\midrule
Geomagnetic field & $+66.6$ & $329.0$ & --- & --- \\
\bottomrule
\end{tabular}
\caption{Natural remanent magnetization directions and moment amplitudes of the five redeposition samples. Rows 2G1--2G4 are the four samples from the first redeposition experiment, measured on a 2G Enterprises superconducting rock magnetometer; 2G5 is the single sample from the second redeposition experiment. Rows QDM1 and QDM2 report the net moment direction obtained from the vector sum of individually inverted dipoles in two QDM fields of view on samples from the first redeposition experiment, following the quality-of-fit filter described in the text. Shallowing is defined as the difference between the applied-field inclination ($+66.6^\circ$) and the measured inclination. Declinations and inclinations are in geographic (true-North) coordinates.}
\label{tab:drm_measurements}
\end{table}

\subsection{Surface magnetometry by Quantum Diamond Microscopy (QDM)}

\begin{figure}[H]
    \centering
    \includegraphics[width=\textwidth]{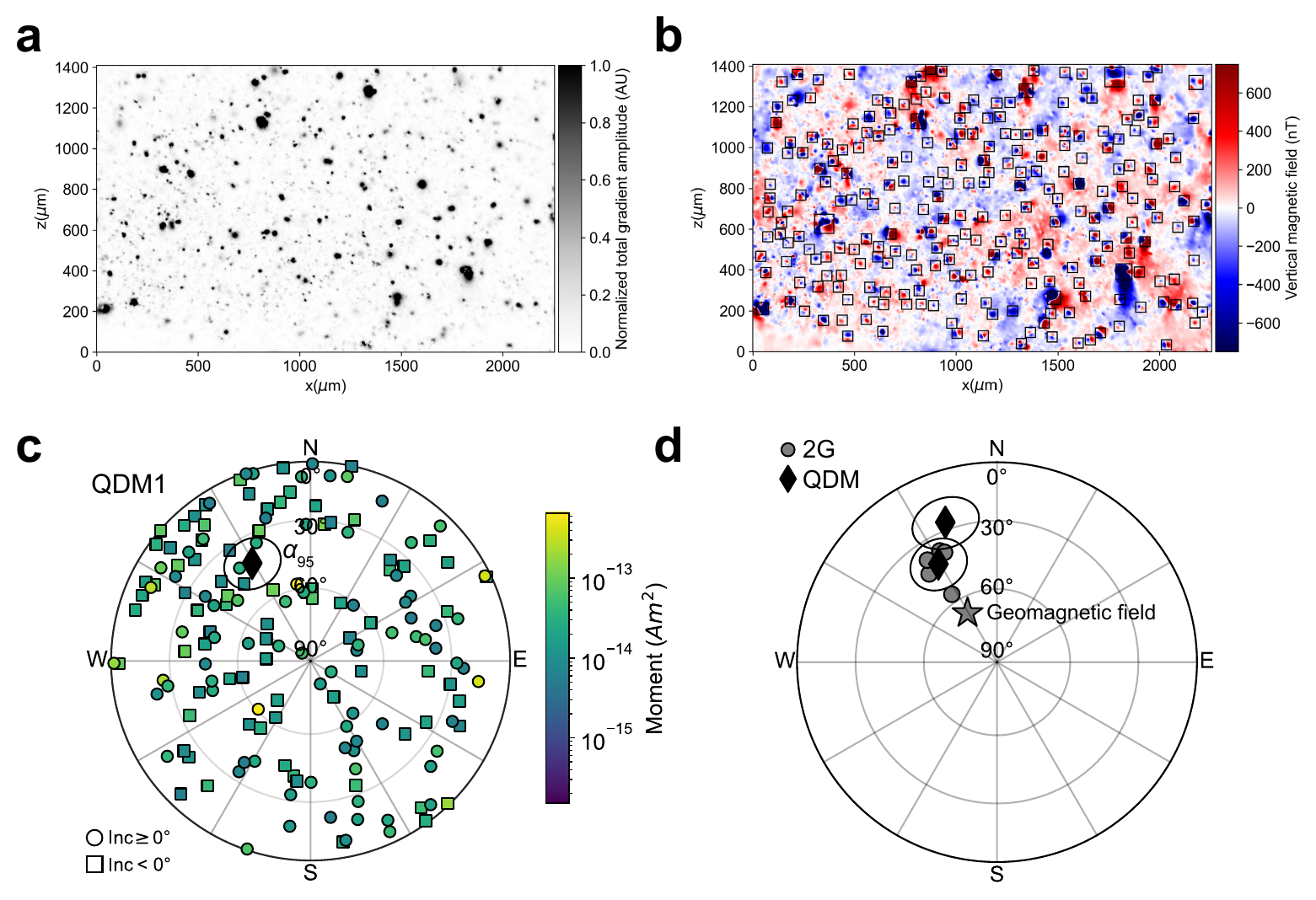}
    \caption{\textbf{Single-particle QDM analysis of a redeposited sediment.}
    \textbf{a}, Normalized total gradient amplitude of the vertical magnetic field over the QDM1 field of view in the sample ($x$, $z$) plane, with $z$ along the water--sediment interface (grayscale, 0 to 1).
    \textbf{b}, QDM map of the vertical magnetic field $B_z$ (nT, red--blue diverging scale) for the same area as panel \textbf{a}, with all $N = 245$ raw dipoles identified by the analysis pipeline in black squares, prior to any filtering.
    \textbf{c}, Equal-area stereographic projection of the directions of the $N = 165$ QDM1 dipoles surviving the dipolarity and moment-amplitude filters (\autoref{tab:qdm_filter_table}): circles denote positive inclinations, squares negative; marker color encodes moment amplitude $|m_i|$. The black diamond marks the net $\%\mathrm{DRM_{moment}}$ direction with its Fisher $\alpha_{95} = 10.96^\circ$ confidence ellipse computed on the moment-weighted (non-unit) vectors.
    \textbf{d}, Equal-area stereographic projection of the five 2G bulk net remanence directions (gray circles), the QDM1 and QDM2 net $\%\mathrm{DRM_{moment}}$ directions (diamonds, with Fisher $\alpha_{95} = 10.96^\circ$ and $11.83^\circ$ confidence ellipses respectively, computed on the moment-weighted vectors), and the applied geomagnetic field at the sampling site (star; inclination $+66.6^\circ$, declination $329.0^\circ$). All inclinations are positive. All measured remanence directions share the applied-field declination within a few degrees and are systematically shallower in inclination.}
    \label{fig:fig2}
\end{figure}

Having established at the bulk level that the redeposition samples carry a strong DRM, we turned to QDM to resolve the directions of individual magnetic carriers. Vector maps of the vertical magnetic field component $B_z$ were acquired across two fields of view on samples from the first redeposition experiment and processed with an anomaly-detection pipeline described in Materials and Methods, which is based on~\cite{SouzaJunior2024}. The pipeline begins with the normalized total gradient amplitude of the measured field (\autoref{fig:fig2}a), which concentrates each source signature on top of its carrier, and returns a catalog of candidate dipoles with their best-fit positions and moment vectors (\autoref{fig:fig2}b). The raw pipeline returned 245 candidate dipoles in the QDM1 field of view. 

We characterize each population using two complementary metrics, defined in \autoref{eq:DRM_definitions}.

\begin{equation}
\label{eq:DRM_definitions}
\mathrm{\%DRM_{moment}}
= 100\,\% \times \frac{\left|\sum_i \mathbf{m}_i\right|}{\sum_i |\mathbf{m}_i|},
\qquad
\mathrm{\%DRM}_{\hat{n}}
= 100\,\% \times \frac{1}{N}\left|\sum_i \frac{\mathbf{m}_i}{|\mathbf{m}_i|}\right|
\end{equation}

$\%\mathrm{DRM_{moment}}$ is the vector sum of dipole moments divided by the sum of their individual magnitudes, which is dominated by the largest-moment carriers and is the direct single-particle analogue of the quantity measured by bulk magnetometry. $\%\mathrm{DRM}_{\hat n}$ is the mean of unit-direction vectors, which weights every detected dipole equally and therefore serves as a directional statistic.

To retain the most reliable particle magnetization estimates, we applied two successive filters (\autoref{tab:qdm_filter_table}): a dipolarity-based quality-of-fit cut ($R^{2} \geq 0.3$, $\mathrm{SNR} \geq 1$) that retains sources well fit by a point-dipole model, and a moment-amplitude cut ($|m_i| \geq 5 \times 10^{-15}$~A\,m$^2$) that restricts the statistical analysis to dipoles above the directional noise of the smallest sources. As we show below, particles below this moment threshold do not contribute strongly to the bulk DRM. The two filters leave $N = 191$ and $N = 165$ dipoles, respectively. The directions of the filtered QDM1 dipoles are shown on a stereonet in \autoref{fig:fig2}c, together with the net $\%\mathrm{DRM_{moment}}$ direction (black diamond) and its Fisher $\alpha_{95}$ confidence cone (black ellipse); the clustering of individual dipoles about the mean is visually apparent. \autoref{fig:fig2}d places the QDM1 and QDM2 net directions alongside the five 2G net vectors and the applied geomagnetic field, showing all remanence directions on a single stereonet. The QDM2 field of view yields 171 dipoles after the same two-stage filter. It reproduces the QDM1 net direction and alignment metrics within experimental error and is reported in full in the Supplementary Information (\autoref{tab:qdm2_filter_table} and \autoref{fig:figure_S1}).

\begin{table}[H]
\centering
\begin{tabular}{lrrr}
\toprule
Quantity & Raw & Dipolarity filter & Dipolarity + $|m|$ filter \\
         &     & ($R^{2} \geq 0.3$, $\mathrm{SNR} \geq 1$) & (plus $|m| \geq 5 \times 10^{-15}$~A\,m$^2$) \\
\midrule
Number of dipoles, $N$                   & 245                   & 191                   & 165                   \\
Inclination ($^\circ$)                   & $+40.2$               & $+42.8$               & $+42.3$               \\
Declination ($^\circ$)                   & $337.6$               & $329.3$               & $329.2$               \\
Net moment amplitude (A\,m$^2$)          & $1.69 \times 10^{-12}$ & $1.80 \times 10^{-12}$ & $1.81 \times 10^{-12}$ \\
\midrule
Watson resultant, $R$                    & $20.50$               & $19.00$               & $23.54$               \\
Watson $R_{\text{crit}}$ ($5\,\%$)       & $25.26$               & $22.31$               & $20.73$               \\
Watson $R_{\text{crit}}$ ($1\,\%$)       & $30.44$               & $26.88$               & $24.98$               \\
Watson test outcome                      & not rejected          & not rejected          & rejected at $5\,\%$   \\
\midrule
$\%\mathrm{DRM_{moment}}$                & $20.3\,\%$            & $23.9\,\%$            & $24.4\,\%$            \\
$\%\mathrm{DRM}_{\hat n}$        & $\phantom{0}8.4\,\%$  & $10.0\,\%$            & $14.3\,\%$            \\
Fisher $\alpha_{95}$ (unit vectors; $^\circ$) & $8.2$            & $9.2$                 & $9.4$                 \\
\bottomrule
\end{tabular}
\caption{QDM1 directional-analysis results under three successive filtering stages. The raw analysis uses all detected dipoles; the second column applies a dipolarity-based quality-of-fit filter requiring $R^{2} \geq 0.3$ and $\mathrm{SNR} \geq 1$; the third additionally restricts to dipoles with inverted moment amplitude $|m_i| \geq 5 \times 10^{-15}$~A\,m$^2$. The net vector inclination and declination are not significantly affected by filtering, indicating that the DRM direction is set primarily by the larger dipoles of the population while the smaller dipoles are more nearly randomly distributed. The Watson randomness test compares the resultant unit-vector length $R$ against the critical values $R_{\text{crit}}$ at the $5\,\%$ and $1\,\%$ significance levels~\cite{Watson1956}; the null hypothesis of an isotropic directional distribution is rejected only after both filters are applied. The moment-weighted and direction-averaged \%DRM metrics follow the definitions in \autoref{eq:DRM_definitions}. Fisher $\alpha_{95}$ is computed on the unit-vector population; the Fisher statistic on the moment-weighted vectors is omitted here because its formulation assumes unit vectors~\cite{fisher1953}.}
\label{tab:qdm_filter_table}
\end{table}

Several features of these measurements are worth noting. The vector sum of the raw $N = 245$ population already defines a net direction (inclination $+40.2^\circ$, declination $337.6^\circ$) within $\sim 9^\circ$ of the applied-field declination and systematically shallower than the ambient $+66.6^\circ$. As the filters are tightened, the net moment amplitude is essentially unchanged ($1.69 \to 1.81 \times 10^{-12}$~A\,m$^2$), indicating that the aligned signal is contained in the subset of highest-moment particles. The declination converges to the applied-field value ($337.6^\circ \to 329.2^\circ$ vs.\ $329.0^\circ$) and the inclination stabilizes at $+42.3^\circ$---within $\sim 3^\circ$ of the mean 2G direction from the same redeposition experiment (\autoref{tab:drm_measurements}) and $\sim 24^\circ$ shallower than the ambient field, reproducing the shallowing observed in bulk magnetometry.

Two results of broad consequence follow. First, the single-particle QDM analysis recovers the direction and inclination shallowing of the bulk 2G measurement from only $\sim 200$ dipoles (\autoref{fig:fig2}d)---orders of magnitude fewer carriers than contribute to a conventional bulk measurement---establishing that DRM can be read out from a sparse (only a few hundred) single-particle population, challenging previous statistical estimates \cite{berndt2016does}. Second, the near-invariance of $\%\mathrm{DRM_{moment}}$ across the filter stages ($20.3\,\% \to 23.9\,\% \to 24.4\,\%$) shows that most of the signal is carried by the highest-moment particles, with the smaller dipoles contributing little additional moment to the DRM sum. 

The direction-averaged metric $\%\mathrm{DRM}_{\hat n}$ weights every carrier equally and is therefore more sensitive to population composition. It is markedly lower than $\%\mathrm{DRM_{moment}}$ ($14.3\,\%$ vs.\ $24.4\,\%$), further indicating that the DRM direction is set by the large-moment carriers while smaller ones are directionally more random, a moment dependence quantified in the particle-level analysis that follows.

\subsection{Particle-level statistical analysis of DRM acquisition}

Access to per-particle moment vectors enables a direct test of which carriers set the DRM direction. For each of the $N = 191$ dipoles surviving the goodness-of-fit filter we compute the alignment $\cos\theta_i = \hat{\mathbf{m}}_i \cdot \hat{\mathbf{n}}_{\text{DRM}}$ with respect to the net $\%\mathrm{DRM_{moment}}$ direction $\hat{\mathbf{n}}_{\text{DRM}}$, and partition the population into five logarithmic bins of moment amplitude spanning $10^{-15}$ to $10^{-12}$~A\,m$^2$ (\autoref{fig:fig3}a). The per-bin mean alignment rises monotonically relative to the population mean of $\langle\cos\theta\rangle = 0.084$: the smallest-moment bin ($N = 29$) sits at $-0.10 \pm 0.10$, consistent within error with directional randomness; the two intermediate bins ($N = 66, 67$) sit at the population mean ($0.08 \pm 0.07$, $0.09 \pm 0.07$); and the two largest-moment bins ($N = 23, 6$) sit at $0.23 \pm 0.09$ and $0.42 \pm 0.19$, factors of $\sim 2.7$ and $\sim 5$ above average. The highest-moment bin should be read cautiously given its small dipole count ($N = 6$), but the trend is already established by the $N = 23$ bin alone.

The cumulative-contribution analysis in \autoref{fig:fig3}b independently confirms this partitioning. Dipoles are added to a running vector sum in order of decreasing $|m|$ (the largest dipoles are added the first) and the cumulative sum magnitude is plotted against the cumulative scalar moment. The first $\sim 20\,\%$ of the total scalar moment---carried by the largest-moment dipoles---produces nearly all of the final vector sum, and the remaining $\sim 80\,\%$ adds essentially no further signal.

Together these analyses establish a size partitioning of DRM acquisition inaccessible to bulk magnetometry: the DRM direction is set almost entirely by a numerical minority of large-moment carriers ($|m| \gtrsim 5 \times 10^{-14}$~A\,m$^2$), while the much more numerous small-moment carriers ($|m| \lesssim 5 \times 10^{-15}$~A\,m$^2$) are directionally random and contribute to neither the direction nor the amplitude of the net remanence. 

This finding has two consequences. First, it provides a particle-level mechanistic basis for the long-standing observation that DRM intensity falls far below the saturation value predicted by equilibrium Langevin alignment~\cite{TauxeSteindorf2006, TauxePaleointensties}: only the subpopulation above a size-dependent alignment threshold contributes meaningfully. Second, it identifies the crossover scale $|m| \sim 10^{-14}$~A\,m$^2$ as the physically relevant window for DRM acquisition---a size threshold above which the field torque dominates thermal randomization.

\subsection{Magnetic mineralogy and domain state}

We performed first-order reversal curve (FORC) and low-temperature magnetization measurements on a sample from the first redeposition experiment (not the one used for the QDM1 and QDM2 measurements) to identify the mineralogy and domain state of the magnetic carriers responsible for the DRM.

The FORC diagram (\autoref{fig:fig4}a) shows a well-defined central ridge with negligible vertical spread, indicating a weakly interacting carrier population spanning the single-domain to single-vortex regime. The coercivity distribution peaks at $\sim 5$--$10$~mT, with the bulk of the intensity below 50~mT and a weak tail extending to $\sim 150$~mT. 

A narrow central ridge of this form is the diagnostic FORC signature of biogenic magnetite produced by magnetotactic bacteria, with chains in varying degrees of integrity — from intact linear chains, through partially bent configurations, to fully collapsed clumps~\cite{egli2010detection, HeslopRoberts2014, pei2023micromagnetic}. Since bending and collapse reduce both chain moment and DRM efficiency~\cite{pei2022bending}, our efficient alignment and large per-particle QDM moments are consistent with chains that are predominantly intact or only weakly bent. The peak position coincides with the assembly-averaged magnetocrystalline coercivity of SD magnetite, and the narrow coercivity range points to near-equant to weakly elongated grains, consistent with magnetosome morphologies of most magnetotactic bacteria~\cite{KoppKirschvink2008}.

Field-cooled and zero-field-cooled remanence curves between 10 and 300~K (\autoref{fig:fig4}b) show a clear Verwey transition between 90 and 120~K, identifying magnetite as the dominant remanence-carrying phase~\cite{Verwey1939, jackson2021distribution}. No secondary transitions attributable to maghemite, hematite, pyrrhotite or greigite are observed.

Together, the FORC signature and the Verwey transition indicate that the DRM is carried by single-domain to single-vortex biogenic magnetite crystals, likely 20--100~nm in size, assembled as magnetosome chains in varying degrees of structural integrity. An individual crystal carries at most $\sim10^{-16}$~Am$^2$, and even an intact magnetosome chain (10-30 crystals) reaches only $\sim10^{-15}$~Am$^2$, which is still one to three orders of magnitude below the dipoles resolved by QDM ($10^{-15}$--$10^{-12}$~Am$^2$). Each QDM dipole must therefore be a composite carrier: a sedimentary floc built from the collapsed or buckled remains of many magnetosome chains, together with associated non-magnetic, fine clastic material. It is these composite carriers, not the individual crystals or intact chains, whose rotation sets the DRM. This is the picture used in the model below.

\subsection{Comparison with an analytical DRM model}

DRM acquisition in natural sediments is a complex, multistage process involving rotational alignment during settling, hydrodynamic torques and inclination shallowing, flocculation, and post-depositional reorganization~\cite{TauxeSteindorf2006, KatariBloxham2001, Heslop2007, LiKodama2016, egli2015natural}. We do not attempt to model this full complexity. Instead, we seek a closed-form description of how alignment depends on carrier moment that can be compared directly with the per-particle QDM measurements, and use it to ask a narrower question: can the magnetic alignment of individual composite carriers account for the measured DRM efficiency on its own, or must additional syn- or post-depositional processes be included?

A natural starting point for DRM acquisition is the equilibrium rotational Langevin model. In this model, non-interacting dipoles reach a Boltzmann distribution in the presence of a magnetic field, such that $\langle\cos\theta\rangle = \mathcal{L}(\alpha) = \coth\alpha - 1/\alpha$, where $\alpha = mB/k_\mathrm{B}T$, $m$ is the magnetic moment, $B$ is the magnetic field strength, $T$ is the temperature, and $k_\mathrm{B}$ is Boltzmann’s constant. Applied to the moments we actually measure by QDM, the model drastically fails. The dipoles in \autoref{fig:fig3} carry $2\times10^{-15}$ to $7\times10^{-13}$~A\,m$^2$ (equivalent to 10 to 10$^5$ single domain grains of magnetite), so at $B = 58.8~\mu$T and the curing temperature of 65\,$^\circ$C, $\alpha$ lies between about 25 and $10^{4}$. Over this range $\mathcal{L}(\alpha)$ is saturated, and the model predicts $\%\mathrm{DRM}\approx99\,\%$ for both metrics instead of the measured 24\,\% and 10\,\%.

The equilibrium assumption is where the model breaks. Because the SD magnetite occurs as inclusions in fine sediment, the composite carriers are flocs of mostly non-magnetic sediment, $\sim$10--50~$\mu$m across (below the 63~$\mu$m sieve size), and their rotation is set by this whole-floc hydrodynamic volume $V_h\approx5\times10^{-16}$--$7\times10^{-14}$~m$^3$. In the EPO-TEK 301 epoxy used here ($\eta_0 = 0.1$--$0.2$~Pa\,s at 23\,$^\circ$C, hardened within $\sim$2~h at 65\,$^\circ$C), thermal rotation at a rate $D_r = k_\mathrm{B}T/6\eta V_h$ would take from half a day to months to randomize such flocs, so they never reach equilibrium (Supplementary Information, \autoref{tab:S-timescales}). Their final orientations are set at dispersion, turned partway toward $\mathbf{B}$ by the magnetic torque and frozen as the resin hardens.

We therefore keep the same non-interacting, overdamped dipoles as magnetic carriers, but follow their rotation in time instead of assuming equilibrium. The difference between the two pictures is clearest if both control parameters are written as the moment divided by a reference moment (Supplementary Information):

\begin{equation}
\alpha = \frac{m}{m_T},\quad m_T = \frac{k_\mathrm{B}T}{B};
\qquad\qquad
\Phi = \frac{m}{m^*},\quad m^* = \frac{6V_h}{B\int\mathrm{d}t/\eta(t)} .
\label{eq:Phi}
\end{equation}

In the equilibrium model, $m_T\approx8\times10^{-17}$~A\,m$^2$ is the moment whose magnetic energy $mB$ equals the thermal energy $k_\mathrm{B}T$, so $\alpha$ compares two \emph{energies}. The kinetic parameter $\Phi$ instead compares two \emph{times}: the time the field needs to rotate a carrier, and the time the carrier is free to rotate before it is locked in place. The first is $\tau_B = 6\eta V_h/mB$, the ratio of rotational drag to magnetic torque, which is the time to turn the carrier through about one radian. $\Phi$ is the second divided by the first (for constant viscosity, $\Phi = t_\mathrm{lock}/\tau_B$). $m^*$ is the moment for which the two times are equal. Carriers with $m\ll m^*$ barely turn before they are locked in, and carriers with $m\gg m^*$ end up fully aligned.

Without thermal noise, the torque balance $6\eta(t)V_h\,\dot\theta = -mB\sin\theta$ integrates to $\tan(\theta/2) = \tan(\theta_0/2)\,e^{-\Phi}$. Averaging over random initial orientations $\theta_0$ (Supplementary Information, Equations S6-S10) gives:

\begin{equation}
\langle\cos\theta\rangle \equiv \mathcal{K}(\Phi) = \coth\Phi - \frac{\Phi}{\sinh^2\Phi}
\label{eq:kinetic}
\end{equation}

$\mathcal{K}$ plays the role of $\mathcal{L}$, and the two functions have similar shapes (\autoref{fig:fig3}c). When alignment is weak both are linear in their argument, $\mathcal{K}\approx2\Phi/3$ and $\mathcal{L}\approx\alpha/3$, so in both models the alignment grows in proportion to $m$ (\autoref{fig:fig3}c). 

What differs is the reference moment. An alignment of $\langle\cos\theta\rangle = 0.1$ requires $m\approx0.3\,m_T\approx2\times10^{-17}$~A\,m$^2$ at equilibrium, but $m\approx0.15\,m^*$ in the kinetic model, which is $\approx6\times10^{-14}$~A\,m$^2$ for the value of $m^*$ found below. That is where our data reach this level of alignment. 


In this kinetic model two torques are neglected. Thermal rotation would lower $\mathcal{K}$ by a factor $1 - m_T/m^*$, less than 0.1\,\% for our system. Hydrodynamic torques are potentially more important, as they can dominate the orientation of non-spherical particles during settling~\cite{Heslop2007}. For a floc (a composite carrier with magnetic and non-magnetic inclusions) whose center of mass is offset from its center of drag by a fraction $\varepsilon$ of its radius $r$, the settling torque is of order $\Delta\rho\,g\,V_h\,\varepsilon r$, independent of viscosity. For 10--50~$\mu$m flocs with $\varepsilon = 0.01$--$0.1$, this exceeds the magnetic torque $mB$ for most of our measured moments, so magnetic alignment is likely effective mainly once settling has largely ceased, in the interval between deposition and lock-in, consistent with the short effective rotation time implied by the fitted $m^*$. Because shape-controlled torques would also yield an alignment that grows in proportion to the moment, our data cannot unambiguously separate them from viscous arrest; $m^*$ is therefore best read as the moment at which the magnetic torque overcomes all non-magnetic constraints on rotation, of which viscous drag is one.

Once $\mathcal{K}$ is known, the population metrics follow as in the equilibrium model, with $\mathcal{L}(\alpha_i)$ replaced by $\mathcal{K}(\Phi_i)$ in the moment-weighted ($\%\mathrm{DRM_{moment}}$) and number-weighted ($\%\mathrm{DRM}_{\hat n}$) averages. 

One more effect matters for a sample of only 191 dipoles. Adding up randomly pointing unit vectors does not give zero. As in a random walk, the sum of $N$ of them has a length of about $\sqrt N$, so randomly oriented dipoles alone would already give $\%\mathrm{DRM}_{\hat n}\approx1/\sqrt N\approx7\,\%$. The same happens for $\%\mathrm{DRM_{moment}}$, but because a few large dipoles dominate the sum, the random baseline is higher, about 14\,\% for our moment distribution. When we predict the two population metrics from the fitted model, we combine this random-walk contribution with the aligned signal (Supplementary Information).

\begin{table}[H]
\centering
\begin{tabular}{rrrrrc}
\toprule
$\langle|m|\rangle$ ($10^{-14}$~A\,m$^2$) & $N$ & $\mathcal{L}(\alpha)$ & $\Phi$ & $\mathcal{K}(\Phi)$ & Measured \\
\midrule
0.37 & 29 & 0.98 & 0.009 & 0.01 & $-0.10\pm0.10$ \\
1.2  & 66 & 0.99 & 0.027 & 0.02 & $\phantom{-}0.08\pm0.07$ \\
3.8  & 67 & 1.00 & 0.088 & 0.06 & $\phantom{-}0.09\pm0.07$ \\
12   & 23 & 1.00 & 0.28  & 0.19 & $\phantom{-}0.23\pm0.09$ \\
39   & 6  & 1.00 & 0.91  & 0.55 & $\phantom{-}0.42\pm0.19$ \\
\midrule
\multicolumn{2}{l}{$\%\mathrm{DRM_{moment}}$} & 99.8 & & 26 & 24 \\
\multicolumn{2}{l}{$\%\mathrm{DRM}_{\hat n}$} & 99.4 & & 10 & 10 \\
\bottomrule
\end{tabular}
\caption{Equilibrium and kinetic predictions compared with the per-bin alignment and the population metrics. $\mathcal{L}(\alpha)$ uses $\alpha = mB/k_\mathrm{B}T$ with $B = 58.8~\mu$T and $T = 338$~K. The kinetic column is $\mathcal{K}(m/m^*)$ with $m^* = 4.3\times10^{-13}$~A\,m$^2$. For the two metrics it includes the random-walk contribution of a finite sample.}
\label{tab:drm_kinetic}
\end{table}

Assuming a single $m^*$ for all dipoles, a weighted fit of \autoref{eq:kinetic} to the five bin means gives $m^* = 4.3^{+1.9}_{-1.1}\times10^{-13}$~A\,m$^2$, with $\chi^2 = 2.8$ for 4 degrees of freedom (\autoref{fig:fig3}d). This fit to the kinetic model produces $\%\mathrm{DRM_{moment}} = 26\,\%$ and $\%\mathrm{DRM}_{\hat n} = 10\,\%$, close to the measured 24\,\% and 10\,\% (\autoref{tab:drm_kinetic}).

The fitted $m^*$ also constrains how the carriers move. Each composite carrier must hold many SD crystals to carry the measured moments (about $10^{4}$ aligned 50~nm crystals for the largest), yet these occupy only $10^{-5}$--$10^{-3}$ of its total floc volume (Supplementary Information). If $m^*$ reflected viscous arrest alone, 50~$\mu$m flocs would satisfy \autoref{eq:Phi} by free rotation in the warm resin, whereas smaller flocs would require $\int\mathrm{d}t/\eta$ up to 40 times smaller. Curing alone cannot supply this, since the integral is set by the early, fluid stage, but the surrounding sediment can: a dense sediment--resin mixture is more viscous than the neat resin, grain contacts hinder rotation further, and shape-controlled torques can also limit alignment. In either case, the DRM falls below the equilibrium prediction because alignment stops well short of equilibrium, the incomplete-equilibration mechanism proposed for natural and laboratory sediments~\cite{TauxeKent2004, TauxeSteindorf2006, KatariBloxham2001}, here resolved at particle level. Since rotation is limited mainly by the surrounding sediment rather than by the resin alone, as it would be in a natural deposit, we expect the alignment and its moment dependence to be representative of natural DRM acquisition, at least in the absence of significant post-depositional disturbance.

\section{Discussion}

Our measurements provide the first per-particle observations of DRM acquisition in the size range most relevant to sedimentary paleomagnetism. Several findings deserve particular emphasis.

First, that a statistically significant DRM direction is recoverable from only $\sim 165$ individually inverted dipole moments---reproducing the bulk 2G direction to within $\sim 3^\circ$ and showing the same $\sim 24^\circ$ inclination shallowing---is itself striking, and the consistency we see across spatially distinct samples establishes that this bias is reproducible rather than idiosyncratic \cite{berndt2016does}. That shallowing arises in a controlled sterile redeposition with minimal post-depositional disturbance adds to evidence that some fraction of DRM shallowing is depositional in origin rather than exclusively compaction-driven~\cite{Tauxe2005, LiKodama2016, PierceSwansonHysell2022}.

Second, the particle-level analysis reveals that DRM depends on magnetic moment: alignment is carried primarily by the largest-moment dipoles, while weaker carriers are directionally random. What matters is moment relative to carrier size, not size itself, because a larger carrier is harder to rotate and more exposed to hydrodynamic torques. This is consistent with the poorer DRM recording of large flocs, in which strong magnetostatic interactions among clustered magnetite crystals further lower the net moment~\cite{egli2010detection, Pei2022}. Importantly, this moment dependence can be resolved directly only by particle-level measurement, and it offers a plausible mechanistic basis for the long-standing DRM efficiency gap: \textit{DRM intensity may be set not by the full carrier population but by the subpopulation above a moment-dependent alignment threshold}. How this scale depends on mineralogy, flocculation and depositional conditions remains to be tested beyond our sample.

Third, the measured DRM is substantial and statistically significant, yet even under sterile conditions designed to minimize post-depositional disturbance it falls well short of the near-complete alignment predicted by equilibrium theory. Both DRM metrics are instead reproduced by a Langevin-based kinetic model in which composite carriers rotate toward the field only for a limited time before the surrounding sediment locks them in place. Incomplete equilibration is therefore not only a consequence of extrinsic randomization but an intrinsic feature of DRM acquisition, present even without mechanical disturbance, compaction or bioturbation, which in natural sediments would lower the recorded remanence further. Because rotation is arrested by the sediment rather than the resin, as in a natural deposit, these conditions are relevant to natural DRM, and longer rotation due to magnetic torques in a less viscous medium, would only strengthen the alignment of large-moment carriers. In this picture alignment efficiency depends on carrier moment, carrier size and the mobility of the host sediment, and should therefore vary between depositional environments, a prediction that particle-level measurements of the kind presented here are well placed to test~\cite{TauxeSteindorf2006, RobertsHuang2020}.


Beyond paleomagnetism, the statistical bias we quantify has direct significance for the prebiotic origin of biological homochirality (\autoref{fig:fig5}). Magnetized magnetite surfaces act as chiral agents through the chiral-induced spin selectivity (CISS) effect, templating enantioselective crystallization of RNA precursors such as ribose aminooxazoline with the preferred handedness set by the surface magnetization direction and reaching enantiomeric excesses of $\sim 60\,\%$ (\autoref{fig:fig5}, left)~\cite{ozturk2023crystallization}. Crystallization of enantiopure precursors can in turn uniformly magnetize initially unmagnetized magnetite, establishing a non-linear feedback between molecular chirality and surface magnetization~\cite{ozturk2023magnetization}. Micromagnetic simulations show this feedback operating on the single-domain to single-vortex domain states characteristic of nanometric magnetite synthesized under prebiotically-plausible pathways~\cite{Devienne2026}. 

This chirality-magnetization feedback can lead to homochirality across a planetary hemisphere only if the magnetic carrier alignment is statistically uniform over scales much larger than the local geochemical environment and stable over prebiotic chemical timescales. Both conditions are satisfied here: the dipolar geomagnetic field imposes a uniform bias far beyond any individual pond or pond system (\autoref{fig:fig5}, right), and single-domain magnetite above the alignment threshold identified here retains its remanence over geological times. A $\sim24\%$ moment-weighted alignment of magnetic dipoles measured at the single-particle level, consistent across independent samples, is therefore wide-scale and long-lived enough to initiate and sustain enantioselective prebiotic chemistry. 

Despite the much weaker geomagnetic field, the natural remanence reaches alignment fractions comparable to the spin polarization produced at synthetic magnetic thin-film surfaces under strong laboratory fields because single-domain inclusions align by free rotation during settling, whereas multi-domain films must be driven across in-plane anisotropy barriers (\autoref{fig:fig5}).

In addition, DRM is not the only natural remanence mechanism observed in depositional environments. Chemical remanent magnetization (CRM) acquired during authigenic production of magnetite in reducing aqueous environments may contribute comparable or greater remanences to the same inventory~\cite{Devienne2026}.

\section{Conclusion}

We have shown that single particle-level quantum diamond microscopy, together with bulk magnetometry, rock-magnetic characterization and a Langevin-based kinetic model, elucidates DRM acquisition at a level inaccessible to bulk magnetometry alone. Alignment is carried predominantly by the largest-moment carriers, consistent with intact magnetosome chains, and the measured DRM is reproduced by composite carriers that rotate toward the field for only a limited time before the surrounding sediment locks them in place. The long-standing DRM efficiency gap therefore need not reflect extrinsic randomization alone: incomplete alignment is intrinsic to DRM acquisition and persists even under sterile, undisturbed conditions. Yet even with incomplete equilibration, the particle-resolved carriers reach a moment-weighted alignment of $\sim$24\,\% of perfect alignment, showing that a substantial DRM can still be achieved when post-depositional disturbance is minimal. This directional bias is statistically uniform across spatially distinct samples and, once locked in, can be preserved over long timescales. 

Our results support that sediments with magnetic carriers, magnetized by the geomagnetic field, can serve as hemispheric-scale and long-lived chiral reagents for initiating and sustaining homochirality through the chiral-induced spin selectivity effect. The DRM framework developed here with single particle-level resolution provides a direct route to investigating these processes across sedimentary settings and to testing naturally magnetized sediments as magnetization-dependent chiral reagents in prebiotic chemistry experiments.

\section*{Acknowledgments}

We thank Alec Brenner for fruitful discussions on bulk magnetometry and for his help in sample polishing. We further thank Joe Kirschvink for reviewing the draft of this manuscript and providing valuable suggestions. S.F.O. acknowledges support from the Kavli-Laukien fellowship program, Caltech startup funds, the William H. Hurt Scholarship Program. R.J.H. and J.A.P.M.D. acknowledge the Leverhulme Centre for Life in the Universe (LCLU) funding (LBAG/329). We thank Cheng Liu for assistance with MPMS measurements that were carried out using Royce equipment at Cambridge Royce facilities under grant EP/P024947/1 and Sir Henry Royce Institute recurrent grant EP/R00661X/1. We also thank Adrian Ioanescu for assistance with VSM measurements in the Department of Materials Science and Metallurgy at the University of Cambridge. 

\subsection*{Author Contributions}

S.F.O., R.R.F., and D.D.S conceived the study. S.F.O. and R.R.F. performed bulk and surface magnetometry experiments and developed the theoretical model. S.F.O. performed data analysis and led the writing of the original draft. J.A.P.M.D. and R.J.H. conducted rock magnetic analyses, and J.A.P.M.D. performed electron microscopy measurements. S.P. and P.T. performed surface magnetometry experiments, and A.T.K. performed bulk magnetometry experiments. S.F.O., R.R.F., J.A.P.M.D., and R.J.H. contributed to manuscript editing. S.F.O., R.R.F., R.J.H., and D.D.S. acquired funding.

\subsection*{Competing Interests}

All authors declare that they have no competing interests.

\subsection*{Data, Code, and Materials Availability}

All data and code needed to evaluate and reproduce the results presented in this paper are provided in the paper and/or the Supplementary Materials. This study did not generate new materials or code.

\newpage

\section*{Materials and Methods}

\subsection*{Reagents and Solvents}
Reagents and solvents were obtained from \textit{Sigma-Aldrich} and \textit{VWR} and were used as received unless otherwise stated.

\subsection*{Sediment Collection and Sieving}
Sediment was collected from two locations in Alewife Brook (Cambridge, MA, USA; 42.40$^\circ$N, 71.15$^\circ$W) using hollow plastic coring tubes. After large debris had been removed, the sediment was transferred to glass containers; the original \emph{in situ} orientation of the cores was not preserved.

The sediment was wet-sieved through a stacked column of USA Standard Sieve Series (ASTM E-11) testing sieves (\textit{Dual Mfg.\ Co.}), with mesh openings of 2~mm, 1~mm, 500~$\mu$m, 250~$\mu$m, 125~$\mu$m, and 63~$\mu$m arranged from coarsest to finest. During sieving, the sediment was rinsed with acetone and ethanol to remove organic matter and soluble material. Only the fraction finer than 63~$\mu$m---that is, the material passing the 63~$\mu$m sieve---was retained for the redeposition experiments.

\subsection*{Redeposition and Epoxy Immobilization}
Throughout redeposition, immobilization, and subsequent specimen preparation, only tools and materials free of magnetic contamination were brought into contact with the sediment, in order not to overprint its remanent magnetization.

A 4~oz clear, straight-sided glass jar (\textit{Specialty Bottle}) was secured with double-sided tape to the hot plate later used for curing (\textit{Corning PC-400D}, 5~$\times$~7~in.\ top, without stirring function). The hot plate was positioned, and the jar fixed to it, at this early stage so that the jar would not need to be moved---and the settled sediment thereby disturbed---at any point before curing. A non-stirring model was chosen deliberately: stirring hot plates drive the stir bar with a motor-driven magnet in the base, the field of which can overprint the geomagnetic remanence of the sample.

A sample coordinate system was defined on the jar. The compass bearing of the sample reference direction (48$^\circ$ east of north) and the local geomagnetic field vector, expressed in sample coordinates as $(x, y, z) = (-12,\ 20,\ -54)$~$\mu$T, were recorded. Together, these quantities allow magnetization directions measured in the magnetometer's instrument coordinates to be transformed into geomagnetic coordinates.

Because the sediment had been disaggregated during sieving, its original depositional remanence was not retained. The sediment was therefore redeposited under controlled conditions so as to acquire a fresh depositional remanent magnetization (DRM) in the known field recorded above. A thin layer (approximately 2~mm) of the sieved sediment was suspended in deionized water in the jar and allowed to settle; as the grains settled, they acquired a DRM aligned with the local geomagnetic field. The layer was kept thin so that, at each subsequent solvent-exchange step, the supernatant could be siphoned from the top and replaced without lifting or pouring from the jar; the shallow depth also ensured that each replacement liquid diffused throughout the full sediment volume.

To immobilize the grains without disturbing the DRM, the pore water was gradually replaced with epoxy resin through a sequence of solvent exchanges. First, the water was exchanged for ethanol through a graded series of water:ethanol mixtures (1:0, 2:1, 1:2, and 0:1 by volume; i.e.\ 0, 33, 67, and 100\% ethanol), each added and withdrawn as described above and with care taken not to disturb the settled sediment. The ethanol was then exchanged for propylene oxide by the same method.

Finally, the propylene oxide was replaced with a graded series of propylene oxide:epoxy-resin mixtures of increasing resin content (1:0, 2:1, 1:2, and 0:1 by volume; i.e.\ 0, 33, 67, and 100\% resin). The epoxy used was \textit{EPO-TEK 301} (\textit{Epoxy Technology}), a low-viscosity (0.1-0.2 Pascal-seconds at room temperature), two-component epoxy that is free of magnetic inclusions and therefore compatible with rock-magnetic measurements; both components were combined before being blended with the propylene oxide. At each step, solvent-rich mixture was withdrawn and resin-rich mixture introduced at the same point using a single-channel air-displacement pipette (\textit{Eppendorf Research plus}, 5~mL), so that any disturbance caused by pipetting was confined to one side of the sample. This side was defined as the 12~o'clock direction in sample coordinates; the diametrically opposite (6~o'clock or the south end in sample coordinates) region remained undisturbed and was reserved for magnetic analysis.

Once infiltration was complete, the hot plate was switched on and the assembly held at 60~$^\circ$C for 48~h to cure the epoxy and ensure complete removal of residual solvent. After curing, the glass jar was broken and the cured sediment--epoxy block was extracted.

\subsection*{Cutting and Polishing}

The block was cut parallel to the direction of gravity from the south end of the sample using acetone-cleaned ceramic knives, exposing the vertical sediment profile. Ceramic rather than metal cutting tools were used to avoid contaminating the sample with magnetic particles. A cyanoacrylate adhesive free of magnetic inclusions (\textit{Krazy Glue}) was drop-cast onto the freshly cut surface to consolidate it, and the sample was placed in a vacuum desiccator to draw out gas bubbles as the adhesive set.

The surface was first ground flat on a lapidary grinder (\textit{Covington Maxi Lap}) and then hand-polished on fine-grit abrasive paper using a figure-eight (``infinity'') motion to minimize directional non-uniformities. Final polishing was performed with an alumina ($\mathrm{Al_2O_3}$) slurry on a polishing pad, down to a grit size of 0.1~$\mu$m, to produce a mirror finish.

\subsection*{Bulk Magnetometry}

Bulk magnetic moment measurements on samples were carried out using a 2G Enterprises model 755 Superconducting Rock Magnetometer (SRM) with moment noise floor of ~$1\times10^{-12}\,Am^2$ and an effective noise floor, limited by the sample holder, of ~$5\times10^{-12}\,Am^2$. The QDM analyzed sample, called SEND4, had an NRM moment of $1.62\times10^{-9}\,Am^2$ and a saturation (1 T) IRM moment of $8.91\times10^{-8}\,Am^2$. The mass of the sample was xx mg, although this value includes a portion of particle-poor epoxy and overestimates the equivalent mass of dry detrital particles by a factor of ~2. 

\subsection*{Surface Magnetometry by a Quantum Diamond Microscope}

We used a quantum diamond microscope, denoted ``Antares'', housed at the Harvard Paleomagnetics Laboratory to acquire all magnetic field maps in this study. Maps were acquired under a 0.9 mT bias field that reversed twice during each integration, resulting in a residual effective bias field up to 400 nT. We corrected for internal reflections of global fluorescence using a factor of 0.2 appropriate for low albedo samples (Fu et al., 2020). Data was taken at 1.17 $\\mu$m per pixel resolution then binned to 2.35 $\\mu$m per pixel for analysis. The polished surface of the SEND4 sample was placed in close contact ($<2$ µm gap) with the QDM for all maps. Raw data was taken in projected magnetic microscopy (PMM) mode along the 111 crystallographic direction of the NV diamond and converted the three-axis readings using the spectral algorithm in Lima and Weiss 2009. 

\subsection*{Quantum Diamond Microscope Data Analysis}

The quantum diamond microscope measurements were first processed in MATLAB to produce a map of the vertical component of the magnetic field, $B_z$, at the measurement plane. This map was exported and used as the input to all subsequent analysis. 

All subsequent analysis was performed in Python using the \textit{micromag} package \cite{SouzaJunior2024} and proceeded in three stages: automatic detection of the magnetic anomaly of each source, inversion of each anomaly for its dipole moment, and directional statistics on the resulting population of dipole moments.

\subsubsection*{Magnetic Anomaly Detection}
The magnetic anomaly associated with each dipolar source was detected and isolated automatically through the following steps:
\begin{enumerate}
    \item A third-order polynomial surface was fitted to the $B_z$ map and subtracted to remove the low-frequency background field.
    
    \item The data were upward-continued by 5~$\mu$m to attenuate high-frequency noise. This continuation height is a map-processing parameter, distinct from the physical sample-to-sensor standoff, and was kept as small as possible so as to suppress noise while retaining the maximum amount of source signal.
    
    \item The total gradient amplitude (TGA) of the upward-continued field was computed to concentrate each anomaly over its source. Spatial derivatives were evaluated using finite differences to avoid amplifying noise.
    
    \item The contrast of the TGA map was stretched (intensity rescaling from 0 to 1) to enhance the visibility of weaker anomalies.
    
    \item Anomalies were located by blob detection on the contrast-stretched TGA map, using the Laplacian-of-Gaussian (LoG) method implemented in \textit{scikit-image} \cite{vanderwalt2014}.
\end{enumerate}

\subsubsection*{Dipole Moment Inversion}
For each detected anomaly, the position of the source was first estimated by Euler deconvolution. The dipole moment was then obtained by linear least-squares inversion of the background-subtracted anomaly data at that fixed position \cite{SouzaJunior2024}. The inversion returns the dipole moment vector together with its covariance matrix, the coefficient of determination of the fit ($R^2$), and the signal-to-noise ratio (SNR) of the anomaly. Each source was thus characterized by its position, dipole moment vector, $R^2$, and SNR.

\subsubsection*{Directional Statistics}
The directions of the fitted dipole moments were analyzed statistically. A test for randomness of directions \cite{Watson1956} was applied to the population of dipole-moment directions to assess whether they are uniformly distributed on the sphere or share a common preferred orientation; the corresponding 95\% and 99\% confidence radii (the 5\% and 1\% significance points) were computed following the procedure of Watson \cite{Watson1956}. The net magnetization direction was obtained as the vector sum of the individual dipole moments, and the semi-angle of its 95\% confidence cone ($\alpha_{95}$) was determined using Fisher statistics \cite{fisher1953}.

This analysis was performed first on the complete set of detected dipoles and then repeated after a quality-control filter that retained only sources with $R^2 > 0.3$ and SNR $> 1$, so that the directional statistics were constrained using only anomalies well described by a dipole model (i.e.\ of high dipolarity). Approximately 200 dipoles were detected within each QDM field of view (FOV) and used to determine the net magnetization direction for that FOV; multiple QDM measurements were acquired and analyzed in this way.

\subsection*{Magnetic Hysteresis and Susceptibility Measurements}
 
Magnetic measurements were carried out on a Lake Shore Cryotronics 8600 vibrating sample magnetometer (VSM) at room temperature. Hysteresis loops were measured in continuous sweep mode up to $\pm1.95$~T with a field step of 10~mT and an averaging time of 0.1~s. An additional loop up to $\pm0.5$~T was measured with 1~mT steps between -100 and +100~mT (10~mT steps outside this range) and an averaging time of 1~s. Hysteresis parameters (saturation magnetization $M_s$, saturation remanence $M_{rs}$ and coercivity $B_c$) were determined after removing the linear high-field (diamagnetic and paramagnetic) contribution. Straight lines were fitted separately to the positive and negative branches at $|B| \geq 1$~T; the two slopes agree within 1\%. The coercivity of remanence $B_{cr}$ was obtained by progressively demagnetizing a saturated sample in a reverse field down to -400~mT. 

Before the FORC measurement the sample was demagnetized in an oscillating, decaying field starting at 1.95~T and 10\% decrement per cycle. We measured 200 FORCs after saturation in +1~T, with reversal fields $B_r$ from +200 to -400~mT and applied fields up to +400~mT, with a field step of 3.015~mT. FORCs were processed using \textit{FORCinel} software (version 3.08) \cite{Harrison2008, egli2013variforc}. Before processing, the lower branch of the hysteresis loop was subtracted from each FORC to generate a set of ``difference FORCs.'' Although this subtraction does not alter the FORC distribution, it substantially reduces common processing artifacts and removes sigmoidal contributions near the remanence diagonal, which cannot be adequately captured by the second-order polynomial function used to smooth the FORC surface \cite{egli2013variforc}. Smoothing factor for FORC processing are S\textsubscript{c0}=8, S\textsubscript{b0}=5, S\textsubscript{c1}=S\textsubscript{b1}=10, $\mathrm{\lambda_h}=\mathrm{\lambda_v}$=0.4.

Low-temperature magnetic susceptibility measurements were performed using a Quantum Design Magnetic Property Measurement System (MPMS-3) at the Maxwell Centre, University of Cambridge. Field cooled (ZFC) curves are obtained by measuring the specimen's magnetic moment in the presence of a DC field (2.5~T) as it cools from room temperature down to $~$2.5 K. Zero-field cooled (ZFC) are also obtained during cooling but in a null magnetic field.

\subsection*{Scanning Electron Microscopy and Energy-Dispersive Spectroscopy}

Scanning electron microscopy (SEM) was performed on a \textit{FEI Quanta 650} FEG-SEM operated in high-vacuum mode. Backscattered-electron (BSE) images were acquired at accelerating voltages of 5--10~kV, and energy-dispersive X-ray spectroscopy (EDS) elemental mapping was performed at 5--15~kV. High-resolution BSE images (1~nm/pixel) were captured using \textit{AZtec} software (version 6.2). The EDS data were analyzed using \textit{HyperSpy} (version 2.3.0) \cite{hyperspy_v2_3_0}. 

\newpage

\printbibliography

@book{dunlop2001rock, 
    place={Cambridge}, 
    series={Cambridge Studies in Magnetism}, 
    title={Rock Magnetism: Fundamentals and Frontiers}, 
    DOI={10.1017/CBO9780511612794}, 
    publisher={Cambridge University Press}, 
    author={Dunlop, David J. and Ozdemir, Ozden}, 
    year={1997}, 
    collection={Cambridge Studies in Magnetism}
}

@article{sallembien2022possible,
  title={Possible chemical and physical scenarios towards biological homochirality},
  author={Sallembien, Quentin and Bouteiller, Laurent and Crassous, Jeanne and Raynal, Matthieu},
  journal={Chemical Society Reviews},
  year={2022},
  publisher={Royal Society of Chemistry}
}

@article{blackmond2019origin,
  title={The origin of biological homochirality},
  author={Blackmond, Donna G},
  journal={cold spring harbor perspectives in biology},
  volume={11},
  number={3},
  pages={a032540},
  year={2019},
  publisher={Cold Spring Harbor Lab}
}

@article{ozturk2023crystallization,
  title={Origin of biological homochirality by crystallization of an RNA precursor on a magnetic surface},
  author={Ozturk, S. Furkan and Liu, Ziwei and Sutherland, John D. and Sasselov, Dimitar D.},
  journal={Science Advances},
  volume={9},
  number={23},
  pages={eadg8274},
  year={2023},
  publisher={American Association for the Advancement of Science}
}

@article{ozturk2023central,
    author = {Ozturk, S. Furkan and Sasselov, Dimitar D. and Sutherland, John D.},
    title = "{The central dogma of biological homochirality: How does chiral information propagate in a prebiotic network?}",
    journal = {The Journal of Chemical Physics},
    volume = {159},
    number = {6},
    pages = {061102},
    year = {2023},
    month = {08},
    issn = {0021-9606},
    doi = {10.1063/5.0156527},
}

@article{ozturk2023magnetization,
  title={Chirality-induced avalanche magnetization of magnetite by an RNA precursor},
  author={Ozturk, S Furkan and Bhowmick, Deb Kumar and Kapon, Yael and Sang, Yutao and Kumar, Anil and Paltiel, Yossi and Naaman, Ron and Sasselov, Dimitar D},
  journal={Nature Communications},
  volume={14},
  number={1},
  pages={6351},
  year={2023},
  publisher={Nature Publishing Group UK London}
}

@article{bloom2024chiral,
  title={Chiral Induced Spin Selectivity},
  author={Bloom, Brian P and Paltiel, Yossi and Naaman, Ron and Waldeck, David H},
  journal={Chemical Reviews},
  year={2024},
  publisher={ACS Publications}
}

@article{Harrison2008,
  author  = {Harrison, R. J. and Feinberg, J. M.},
  title   = {FORCinel: An improved algorithm for calculating first-order reversal curve distributions using locally weighted regression smoothing},
  journal = {Geochemistry, Geophysics, Geosystems},
  volume  = {9},
  pages   = {Q05016},
  year    = {2008}
}

@article{egli2013variforc,
  title={VARIFORC: An optimized protocol for calculating non-regular first-order reversal curve (FORC) diagrams},
  author={Egli, R},
  journal={Global and Planetary Change},
  volume={110},
  pages={302--320},
  year={2013},
  publisher={Elsevier}
}

@article{pei2022bending,
  title={Bending and collapse: Magnetic recording fidelity of magnetofossils from micromagnetic simulation},
  author={Pei, Zhaowen and Berndt, Thomas A and Chang, Liao and Bai, Fan and Williams, Wyn and Paterson, Greig A},
  journal={Journal of Geophysical Research: Solid Earth},
  volume={127},
  number={4},
  pages={e2021JB023447},
  year={2022},
  publisher={Wiley Online Library}
}

@article{kaub2026magnetic,
  title={Magnetic dipole imaging of magnetite nanoparticles in brain tissue},
  author={Kaub, Leon and Gilder, Stuart A and Fu, Roger R and Maher, Barbara A and Maxemin, Gabriel and Kuan, Aaron T and B{\"u}ttner, Andreas and Milz, Stefan and Schmitz, Christoph},
  journal={RSC advances},
  volume={16},
  number={2},
  pages={983--994},
  year={2026},
  publisher={The Royal Society of Chemistry}
}

@article{roberts2013magnetic,
  title={Magnetic paleointensity stratigraphy and high-resolution Quaternary geochronology: successes and future challenges},
  author={Roberts, Andrew P and Tauxe, Lisa and Heslop, David},
  journal={Quaternary Science Reviews},
  volume={61},
  pages={1--16},
  year={2013},
  publisher={Elsevier}
}

@article{jackson2021distribution,
  author  = {Jackson, Mike J. and Moskowitz, Bruce},
  title   = {On the distribution of Verwey transition temperatures in natural magnetites},
  journal = {Geophysical Journal International},
  volume  = {224},
  number  = {2},
  pages   = {1314--1325},
  month   = feb,
  year    = {2021},
  doi     = {10.1093/gji/ggaa516},
  url     = {https://doi.org/10.1093/gji/ggaa516}
}

@article{Dunlop2002DayPlot,
  author    = {Dunlop, D. J.},
  title     = {Theory and application of the Day plot (Mrs/Ms versus Hcr/Hc), 1. Theoretical curves and tests using titanomagnetite data},
  journal   = {Journal of Geophysical Research},
  volume    = {107},
  number    = {B3},
  year      = {2002},
  doi       = {10.1029/2001JB000486}
}

@article{Day1977Hysteresis,
  author    = {Day, R. and Fuller, M. and Schmidt, V. A.},
  title     = {Hysteresis properties of titanomagnetites: Grain size and composition dependence},
  journal   = {Physics of the Earth and Planetary Interiors},
  volume    = {13},
  pages     = {260--267},
  year      = {1977}
}

@article{Ozturk2025Homochirality,
  author    = {Ozturk, S. F. and Sasselov, D. D.},
  title     = {Life's homochirality: Across a prebiotic network},
  journal   = {Proceedings of the National Academy of Sciences of the United States of America},
  volume    = {122},
  number    = {34},
  pages     = {e2505126122},
  year      = {2025},
  doi       = {10.1073/pnas.2505126122}
}

@software{hyperspy_v2_3_0,
  author       = {de la Pe{\~n}a, Francisco and
                  Prestat, Eric and
                  L{\"a}hnemann, Jonas and
                  Fauske, Vidar Tonaas and
                  Burdet, Pierre and
                  Jokubauskas, Petras and
                  Furnival, Tom and
                  Francis, Carter and
                  Nord, Magnus and
                  Ostasevicius, Tomas and
                  MacArthur, Katherine E. and
                  Johnstone, Duncan N. and
                  Sarahan, Mike and
                  Taillon, Joshua and
                  Aarholt, Thomas and
                  Quinn, Patrick and
                  Migunov, Vadim and
                  Eljarrat, Alberto and
                  Caron, Jan and
                  Nemoto, Takeshi and
                  Poon, Tim and
                  Mazzucco, Stefano and
                  Borg, Siv and
                  Tappy, Nicolas and
                  Cautaerts, Niels and
                  Somnath, Suhas and
                  Slater, Tom and
                  Walls, Matthew and
                  pietsjoh},
  title        = {{hyperspy/hyperspy: v2.3.0}},
  version      = {v2.3.0},
  year         = {2025},
  publisher    = {Zenodo},
  doi          = {10.5281/zenodo.14956374},
  url          = {https://doi.org/10.5281/zenodo.14956374}
}

@article{vanderwalt2014,
  author  = {van der Walt, St{\'e}fan and Sch{\"o}nberger, Johannes L. and Nunez-Iglesias, Juan and Boulogne, Fran{\c{c}}ois and Warner, Joshua D. and Yager, Neil and Gouillart, Emmanuelle and Yu, Tony and the scikit-image contributors},
  title   = {scikit-image: image processing in {Python}},
  journal = {PeerJ},
  volume  = {2},
  pages   = {e453},
  year    = {2014},
  doi     = {10.7717/peerj.453}
}

@article{fisher1953,
  author  = {Fisher, Ronald A.},
  title   = {Dispersion on a Sphere},
  journal = {Proceedings of the Royal Society of London. Series A},
  volume  = {217},
  number  = {1130},
  pages   = {295--305},
  year    = {1953},
  doi     = {10.1098/rspa.1953.0064}
}

@book{TauxeBook,
  author    = {Tauxe, Lisa},
  title     = {Essentials of Paleomagnetism},
  publisher = {University of California Press},
  address   = {Berkeley, California},
  year      = {2010}
}

@book{nagata1961rock,
  title={Rock magnetism},
  author={Nagata, Takeshi},
  year={1961},
  publisher={Maruzen Company}
}

@article{Tauxe1993,
  author  = {Tauxe, Lisa},
  title   = {Sedimentary records of relative paleointensity of the geomagnetic field: theory and practice},
  journal = {Reviews of Geophysics},
  volume  = {31},
  number  = {3},
  pages   = {319--354},
  year    = {1993},
  doi     = {10.1029/93RG01771}
}

@incollection{TauxePaleointensties,
title = {5.13 - Paleointensities},
editor = {Gerald Schubert},
booktitle = {Treatise on Geophysics (Second Edition)},
publisher = {Elsevier},
edition = {Second Edition},
address = {Oxford},
pages = {461-509},
year = {2015},
isbn = {978-0-444-53803-1},
doi = {https://doi.org/10.1016/B978-0-444-53802-4.00107-X},
url = {https://www.sciencedirect.com/science/article/pii/B978044453802400107X},
author = {L. Tauxe and T. Yamazaki},
}

@incollection{TauxeKent2004,
  author    = {Tauxe, Lisa and Kent, Dennis V.},
  title     = {A simplified statistical model for the geomagnetic field and the detection of shallow bias in paleomagnetic inclinations: Was the ancient magnetic field dipolar?},
  booktitle = {Timescales of the Paleomagnetic Field},
  editor    = {Channell, James E. T. and Kent, Dennis V. and Lowrie, William and Meert, Joseph G.},
  series    = {Geophysical Monograph},
  volume    = {145},
  pages     = {101--115},
  publisher = {American Geophysical Union},
  address   = {Washington, D.C.},
  year      = {2004},
  doi       = {10.1029/145GM08}
}

@article{Tauxe2005,
  author  = {Tauxe, Lisa},
  title   = {Inclination flattening and the geocentric axial dipole hypothesis},
  journal = {Earth and Planetary Science Letters},
  volume  = {233},
  pages   = {247--261},
  year    = {2005},
  doi     = {10.1016/j.epsl.2005.01.027}
}

@article{TauxeSteindorf2006,
  author  = {Tauxe, Lisa and Steindorf, Jason L. and Harris, Andrew},
  title   = {Depositional remanent magnetization: Toward an improved theoretical and experimental foundation},
  journal = {Earth and Planetary Science Letters},
  volume  = {244},
  number  = {3--4},
  pages   = {515--529},
  year    = {2006},
  doi     = {10.1016/j.epsl.2006.02.003}
}

@article{Heslop2007,
  author  = {Heslop, David},
  title   = {Are hydrodynamic shape effects important when modelling the formation of depositional remanent magnetization?},
  journal = {Geophysical Journal International},
  volume  = {171},
  number  = {3},
  pages   = {1029--1035},
  year    = {2007},
  doi     = {10.1111/j.1365-246X.2007.03588.x}
}

@article{LiKodama2016,
  author  = {Li, Youjun and Kodama, Kenneth P.},
  title   = {Detecting and correcting for paleomagnetic inclination shallowing of sedimentary rocks: A review},
  journal = {Frontiers in Earth Science},
  volume  = {4},
  pages   = {7},
  year    = {2016},
  doi     = {10.3389/feart.2016.00007}
}

@article{MitraTauxe2009,
  author  = {Mitra, Ritayan and Tauxe, Lisa},
  title   = {Full vector model for magnetization in sediments},
  journal = {Earth and Planetary Science Letters},
  volume  = {286},
  number  = {3--4},
  pages   = {535--545},
  year    = {2009},
  doi     = {10.1016/j.epsl.2009.07.009}
}

@article{RobertsHuang2020,
  author  = {Roberts, Andrew P. and Zhao, Xiang and Hu, Pengxiang and Abrajevitch, Alexandra and Chen, Yen-Hua and Harrison, Richard J. and Heslop, David and Jiang, Zhaoxia and Li, Jinhua and Liu, Qingsong and Muxworthy, Adrian R. and Oda, Hirokuni and Sato, Tetsuro and Tauxe, Lisa and Zhang, Weilin},
  title   = {Detrital remanent magnetization of single-crystal silicates with magnetic inclusions: constraints from deposition experiments},
  journal = {Geophysical Journal International},
  volume  = {224},
  number  = {3},
  pages   = {2001--2020},
  year    = {2021},
  doi     = {10.1093/gji/ggaa559}
}

@article{anson1987compaction,
  title={Compaction-induced inclination shallowing of the post-depositional remanent magnetization in a synthetic sediment},
  author={Anson, GL and Kodama, KP},
  journal={Geophysical Journal International},
  volume={88},
  number={3},
  pages={673--692},
  year={1987},
  publisher={Blackwell Publishing Ltd Oxford, UK}
}

@article{ShcherbakovSycheva2010,
  author  = {Shcherbakov, Valera P. and Sycheva, Natalia K.},
  title   = {On the mechanism of formation of depositional remanent magnetization},
  journal = {Geochemistry, Geophysics, Geosystems},
  volume  = {11},
  number  = {2},
  pages   = {Q02Z13},
  year    = {2010},
  doi     = {10.1029/2009GC002830}
}

@article{King1955,
  author  = {King, Robert F.},
  title   = {The remanent magnetism of artificially deposited sediments},
  journal = {Monthly Notices of the Royal Astronomical Society, Geophysical Supplement},
  volume  = {7},
  number  = {3},
  pages   = {115--134},
  year    = {1955},
  doi     = {10.1111/j.1365-246X.1955.tb06558.x}
}

@article{TauxeKent1984,
  author  = {Tauxe, Lisa and Kent, Dennis V.},
  title   = {Properties of a detrital remanence carried by haematite from study of modern river deposits and laboratory redeposition experiments},
  journal = {Geophysical Journal of the Royal Astronomical Society},
  volume  = {76},
  number  = {3},
  pages   = {543--561},
  year    = {1984},
  doi     = {10.1111/j.1365-246X.1984.tb01909.x}
}

@article{Bilardello2013,
  author  = {Bilardello, Dario and Jackson, Mike J.},
  title   = {What do the {Mumpsimus} diagrams tell us about long-term inclination shallowing? Some ideas from {FORC} and low-{T} experiments},
  journal = {Geophysical Journal International},
  volume  = {195},
  number  = {1},
  pages   = {148--163},
  year    = {2013},
  doi     = {10.1093/gji/ggt254}
}

@article{PierceSwansonHysell2022,
  author  = {Pierce, Elizabeth L. and Swanson-Hysell, Nicholas L.},
  title   = {Quantifying inclination shallowing and representing flattening uncertainty in sedimentary paleomagnetic poles},
  journal = {Geochemistry, Geophysics, Geosystems},
  volume  = {23},
  number  = {11},
  pages   = {e2022GC010682},
  year    = {2022},
  doi     = {10.1029/2022GC010682}
}

@article{egli2010detection,
  title={Detection of noninteracting single domain particles using first-order reversal curve diagrams},
  author={Egli, Ramon and Chen, Amy P and Winklhofer, Michael and Kodama, Kenneth P and Horng, Chorng-Shern},
  journal={Geochemistry, Geophysics, Geosystems},
  volume={11},
  number={1},
  year={2010},
  publisher={Wiley Online Library}
}

@article{egli2015natural,
  title={Natural remanent magnetization acquisition in bioturbated sediment: General theory and implications for relative paleointensity reconstructions},
  author={Egli, R and Zhao, X},
  journal={Geochemistry, Geophysics, Geosystems},
  volume={16},
  number={4},
  pages={995--1016},
  year={2015},
  publisher={Wiley Online Library}
}

@article{pei2023micromagnetic,
  title={Micromagnetic calculation of the magnetite magnetosomal morphology control of magnetism in magnetotactic bacteria},
  author={Pei, Zhaowen and Chang, Liao and Bai, Fan and Harrison, Richard J},
  journal={Journal of the Royal Society Interface},
  volume={20},
  number={206},
  pages={20230297},
  year={2023}
}

@article{HeslopRoberts2014,
  author  = {Heslop, David and Roberts, Andrew P. and Chang, Liao and Davies, Maureen and Abrajevitch, Alexandra and De Deckker, Patrick},
  title   = {Quantifying magnetite magnetofossil contributions to sedimentary magnetizations},
  journal = {Earth and Planetary Science Letters},
  volume  = {382},
  pages   = {58--65},
  year    = {2013},
  doi     = {10.1016/j.epsl.2013.09.011}
}

@article{KoppKirschvink2008,
  author  = {Kopp, Robert E. and Kirschvink, Joseph L.},
  title   = {The identification and biogeochemical interpretation of fossil magnetotactic bacteria},
  journal = {Earth-Science Reviews},
  volume  = {86},
  number  = {1--4},
  pages   = {42--61},
  year    = {2008},
  doi     = {10.1016/j.earscirev.2007.08.001}
}

@article{GlennFu2017,
  author  = {Glenn, David R. and Fu, Roger R. and Kehayias, Pauli and Le Sage, David and Lima, Eduardo A. and Weiss, Benjamin P. and Walsworth, Ronald L.},
  title   = {Micrometer-scale magnetic imaging of geological samples using a quantum diamond microscope},
  journal = {Geochemistry, Geophysics, Geosystems},
  volume  = {18},
  number  = {8},
  pages   = {3254--3267},
  year    = {2017},
  doi     = {10.1002/2017GC006946}
}

@article{QDMlab2022,
title = {QDMlab: A MATLAB toolbox for analyzing quantum diamond microscope (QDM) magnetic field maps},
journal = {Computers \& Geosciences},
volume = {167},
pages = {105198},
year = {2022},
issn = {0098-3004},
doi = {https://doi.org/10.1016/j.cageo.2022.105198},
author = {Michael W.R. Volk and Roger R. Fu and Raisa Trubko and Pauli Kehayias and David R. Glenn and Eduardo A. Lima}
}

@article{SouzaJunior2024,
  author  = {Souza-Junior, Gelson F. and Uieda, Leonardo and Trindade, Ricardo I. F. and Carmo, Janine and Fu, Roger R.},
  title   = {Full vector inversion of magnetic microscopy images using {Euler} deconvolution as prior information},
  journal = {Geochemistry, Geophysics, Geosystems},
  volume  = {25},
  number  = {3},
  pages   = {e2023GC011082},
  year    = {2024},
  doi     = {10.1029/2023GC011082}
}

@article{Verwey1939,
  author  = {Verwey, Evert J. W.},
  title   = {Electronic conduction of magnetite ({Fe$_3$O$_4$}) and its transition point at low temperatures},
  journal = {Nature},
  volume  = {144},
  number  = {3642},
  pages   = {327--328},
  year    = {1939},
  doi     = {10.1038/144327b0}
}

@article{Watson1956,
  author  = {Watson, Geoffrey S.},
  title   = {A test for randomness of directions},
  journal = {Monthly Notices of the Royal Astronomical Society, Geophysical Supplement},
  volume  = {7},
  number  = {4},
  pages   = {160--161},
  year    = {1956},
  doi     = {10.1111/j.1365-246X.1956.tb05561.x}
}

@article{Devienne2026,
author = {José A. P. M. Devienne  and Ziwei Liu  and Clancy Zhijian Jiang  and Nicholas J. Tosca  and Thomas Ginnis  and Dimitar D. Sasselov  and Richard J. Harrison  and S. Furkan Ozturk },
title = {Prebiotic magnetite enables chirality-magnetic surface feedback},
journal = {Science Advances},
volume = {12},
number = {45},
pages = {eaeg3954},
year = {2026},
doi = {10.1126/sciadv.aeg3954},
}

@article{KatariBloxham2001,
  author  = {Katari, Kartik and Bloxham, Jeremy},
  title   = {Effects of sediment aggregate size on {DRM} intensity: a new theory},
  journal = {Earth and Planetary Science Letters},
  volume  = {186},
  number  = {1},
  pages   = {113--122},
  year    = {2001}
}

@article{Pei2022,
  author  = {Pei, Zhaowen and Berndt, Thomas A. and Chang, Liao and Bai, Fan and Williams, Wyn and Paterson, Greig A.},
  title   = {Bending and collapse: Magnetic recording fidelity of magnetofossils from micromagnetic simulation},
  journal = {Journal of Geophysical Research: Solid Earth},
  volume  = {127},
  number  = {4},
  pages   = {e2021JB023447},
  year    = {2022},
  doi     = {10.1029/2021JB023447}
}

@article{piascik2025reliable,
  title={Reliable identification of flood-borne deposits in speleothems using magnetic microscopy},
  author={Piascik, Samuel L and Fu, Roger R and Str{\'\i}kis, Nicol{\'a}s M and Trindade, Ricardo IF and Rodrigues, Julio Cauhy and Cruz, Francisco W and Souza-Junior, Gelson F and Borsato, Andrea and Azevedo, Vitor and Jaqueto, Plinio and others},
  journal={Quaternary Science Reviews},
  volume={369},
  pages={109638},
  year={2025},
  publisher={Elsevier}
}

@article{bellon2026estimating,
  title={Estimating Earth’s past field strength from individual sources in archaeological ceramics using quantum diamond microscopy},
  author={Bellon, Ualisson D and F Souza-Junior, Gelson and Williams, Wyn and Muxworthy, Adrian R and Uieda, Leonardo and Fu, Roger R and Trindade, Ricardo IF and Moraes, Carolina S},
  journal={Communications Earth \& Environment},
  volume={7},
  number={1},
  pages={766},
  year={2026},
  publisher={Nature Publishing Group}
}

@article{qi2025quantifying,
  author  = {Qi, L. and Muxworthy, A. R. and Baker, E. B. and Cao, X. and Allerton, S. and Bryson, J. F. J. and Zhang, Y.},
  title   = {Quantifying serpentinization-driven remagnetization from ridge axis to subduction zone using quantum diamond microscopy},
  journal = {Journal of Geophysical Research: Solid Earth},
  volume  = {130},
  pages   = {e2025JB031606},
  year    = {2025},
  doi     = {10.1029/2025JB031606}
}

@article{fu2020high,
  title={High-sensitivity moment magnetometry with the quantum diamond microscope},
  author={Fu, Roger R and Lima, Eduardo A and Volk, Michael WR and Trubko, Raisa},
  journal={Geochemistry, Geophysics, Geosystems},
  volume={21},
  number={8},
  pages={e2020GC009147},
  year={2020},
  publisher={Wiley Online Library}
}

@article{berndt2016does,
  title={Does size matter? Statistical limits of paleomagnetic field reconstruction from small rock specimens},
  author={Berndt, Thomas and Muxworthy, Adrian R and Fabian, Karl},
  journal={Journal of Geophysical Research: Solid Earth},
  volume={121},
  number={1},
  pages={15--26},
  year={2016},
  publisher={Wiley Online Library}
}

\newpage

\newpage

\newpage

\begin{figure}[H]
    \centering
    \includegraphics[width=\textwidth]{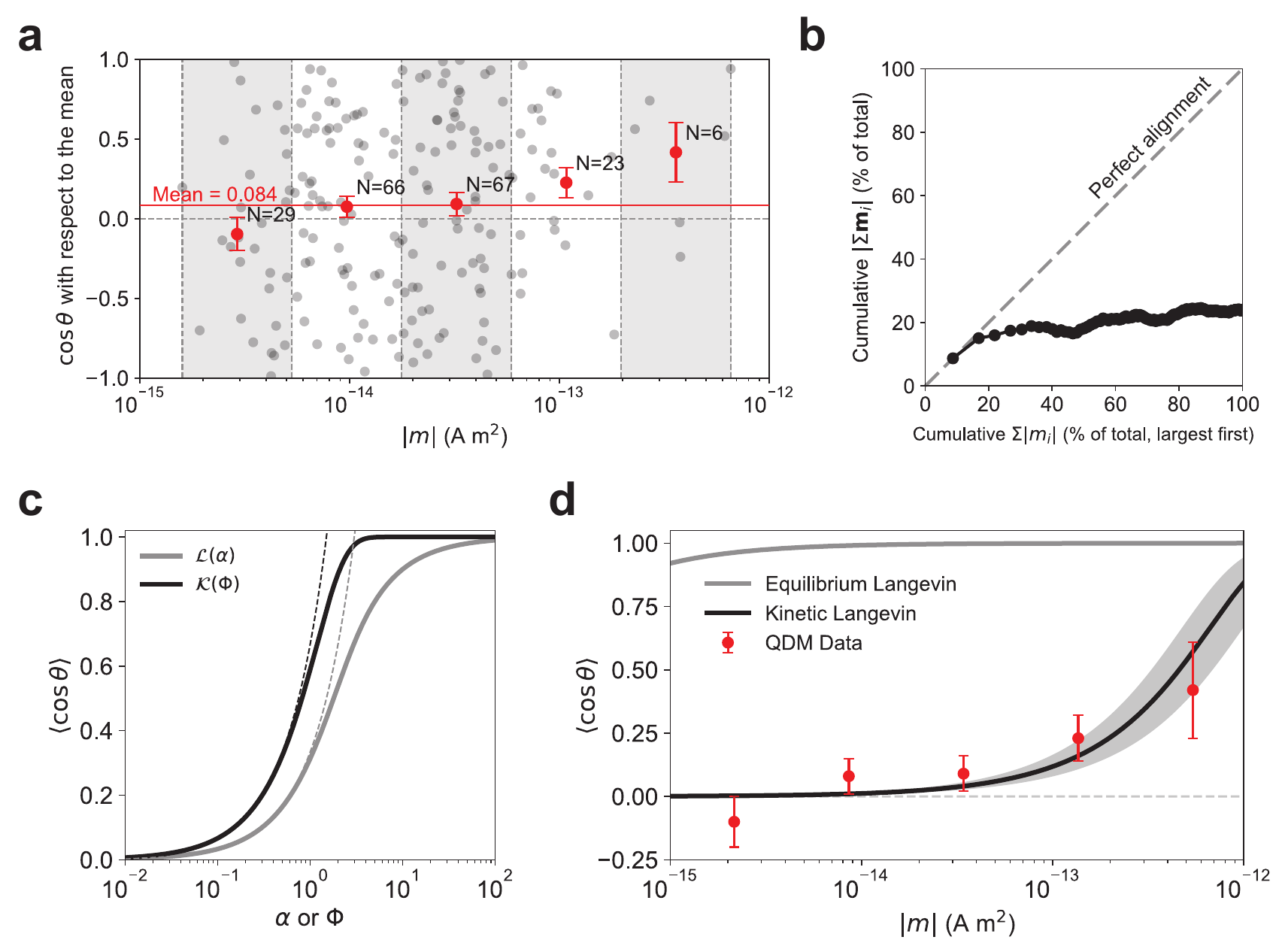}
    \caption{\textbf{Particle-level statistical analysis of DRM and kinetic alignment model.} \textbf{a}, Per-dipole alignment with respect to the net $\%\mathrm{DRM_{moment}}$ direction as a function of moment amplitude $|m_i|$ for the $N = 191$ dipoles surviving the dipolarity filter (gray disks). Red symbols with error bars are per-bin means ($\pm$ standard error of the mean) in five logarithmic $|m|$ bins; bin counts $N$ are annotated. The horizontal red line marks the population mean, $\langle\cos\theta\rangle = 0.084$. \textbf{b}, Cumulative-contribution plot: dipoles are added to a running vector sum in order of decreasing $|m|$, and the cumulative sum magnitude $|\sum\mathbf{m}_i|$ is plotted against the cumulative scalar moment $\sum|m_i|$, both normalized to their final totals. The dashed diagonal marks perfect alignment of all constituent moments. The data curve reaches close to its final value after only $\sim20\,\%$ of the total scalar moment is added, showing that the largest-moment dipoles carry most of the aligned DRM signal. \textbf{c}, Mean alignment predicted by the equilibrium Langevin function, $\mathcal{L}(\alpha)$ with $\alpha = mB/k_\mathrm{B}T$ (gray), and by the kinetic Langevin function, $\mathcal{K}(\Phi)$ with $\Phi = m/m^*$ (black), for carriers that rotate toward the field until the resin arrests them. Dashed lines are the weak-alignment limits $\alpha/3$ and $2\Phi/3$. \textbf{d}, Per-bin mean alignment from \textbf{a} compared with the equilibrium Langevin prediction $\mathcal{L}(mB/k_\mathrm{B}T)$ (gray) and the kinetic model $\mathcal{K}(m/m^*)$ fitted to the bin means (black), with $m^* = 4.3\times10^{-13}$~A\,m$^2$. The shaded band spans the 68\,\% confidence range of $m^*$. The equilibrium model predicts near-complete alignment for all measured moments, whereas the kinetic model reproduces the weak, moment-dependent alignment.}
    \label{fig:fig3}
\end{figure}

\newpage


\begin{figure}[H]
    \centering
    \includegraphics[width=\textwidth]{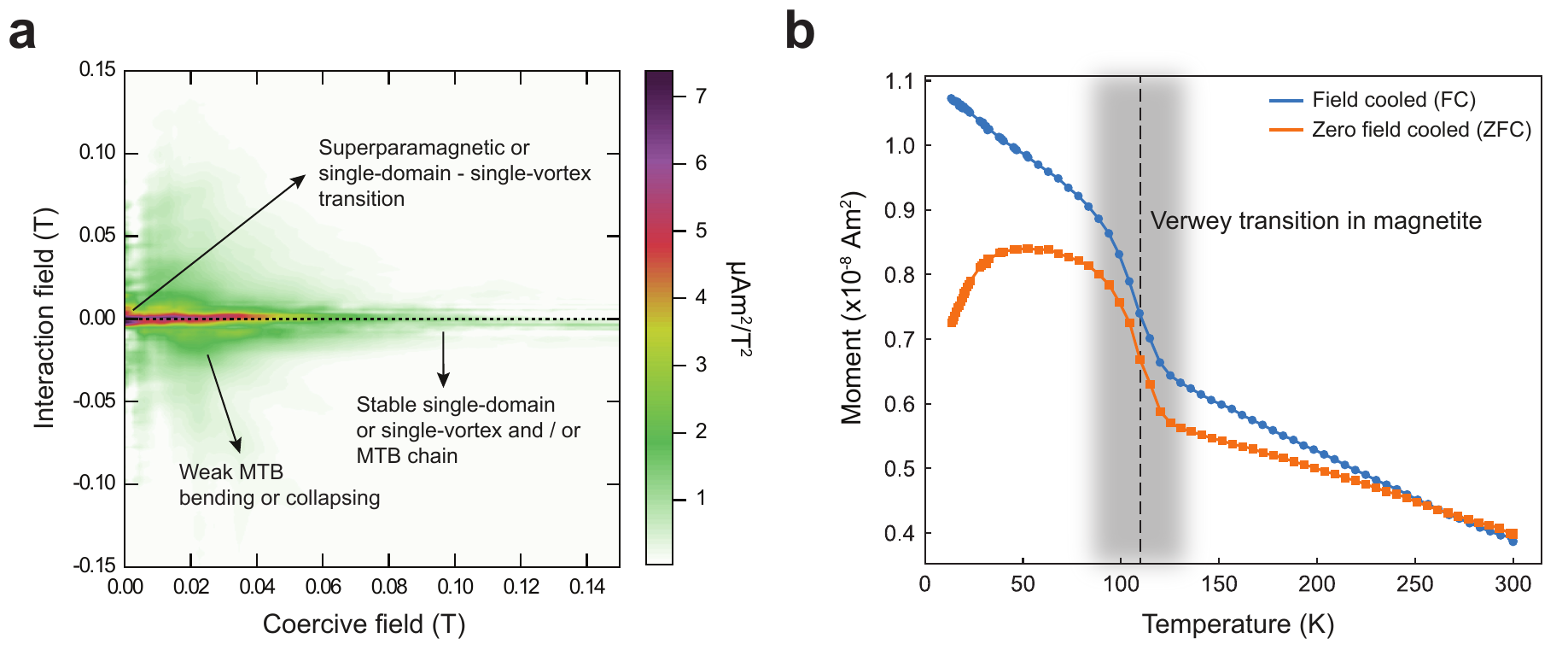}
    \caption{\textbf{Magnetic mineralogy and domain state of the redeposited sample.} \textbf{a}, First-order reversal curve (FORC) diagram of a sample from the first redeposition experiment, plotted in coercive field ($B_c$) and interaction field ($B_u$) coordinates, with FORC density on a yellow--green colour scale ($\mu$A\,m$^2$/T$^2$). The narrow central ridge at $B_u \approx 0$, with coercivity peaking at $\sim 5$--$10$~mT, indicates weakly interacting single-domain to single-vortex grains and is characteristic of biogenic magnetite from magnetotactic bacteria (MTB) species. \textbf{b}, Field-cooled (FC, blue) and zero-field-cooled (ZFC, orange) remanence from 10 to 300~K. The drop in both curves between $\sim 90$ and $\sim 120$~K (gray shaded region) is the Verwey transition, identifying magnetite as the dominant remanence carrier.}
    \label{fig:fig4}
\end{figure}

\newpage

\begin{figure}[H]
    \centering
    \includegraphics[width=\textwidth]{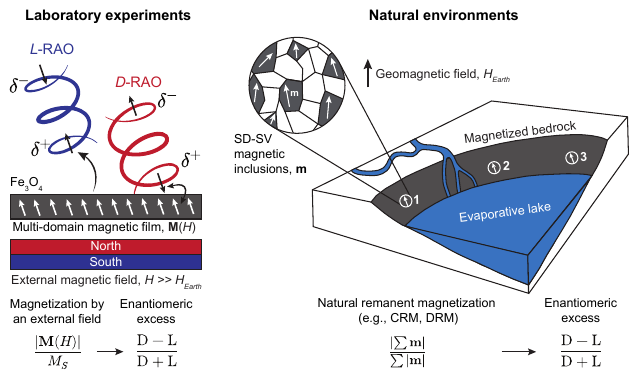}
    \caption{\textbf{Natural remanent magnetization as a chiral agent for prebiotic chemistry.}
    \textbf{Left:} laboratory experiment. A multi-domain magnetite (Fe$_3$O$_4$) thin film is magnetized by an external field $H$, producing a surface magnetization $\mathbf{M}(H)$. Through the chiral-induced spin selectivity effect, its normalized amplitude $|\mathbf{M}(H)|/M_s$ sets the magnitude of the enantiomeric excess $(D-L)/(D+L)$ in the crystallization of chiral molecules---here the RNA precursor ribose aminooxazoline (RAO) in its $L$ (blue) and $D$ (red) enantiomers---while the sign of $\mathbf{M}$ selects which enantiomer is preferred.
    \textbf{Right:} natural analog. Lacustrine sediments of an evaporative lake acquire a natural remanent magnetization (NRM) in the ambient geomagnetic field $H$\textsubscript{Earth}, carried by single-domain (SD) to single-vortex (SV) magnetic inclusions ($\mathbf{m}$) within the sedimentary host (inset). The field-aligned carrier fraction $|\sum \mathbf{m}|/\sum|\mathbf{m}|$, quantified in this work, is the natural counterpart of $|\mathbf{M}(H)|/M_s$ in the laboratory analog. It remains comparable in magnitude despite the much weaker aligning field ($H$\textsubscript{Earth}$\ll H$) because fine SD inclusions reach substantial alignment during settling, whereas multi-domain thin films must be driven across in-plane anisotropy barriers. The geomagnetic dipole ensures that this alignment is spatially coherent across scales (locations 1--3) that far exceed any individual pond. Together with the spin-controlled enantioselectivity of magnetite surfaces, this spatially uniform, long-lived remanence provides a deterministic route to initiating and sustaining biomolecular homochirality on early Earth.}
    \label{fig:fig5}  
\end{figure}

\newpage

\begin{center}

{\Large\bfseries Supplementary Information\par}
\vspace{0.5em}

{\large\bfseries Detrital Remanent Magnetization Analysis at the Particle Scale: Implications for Prebiotic Origins of Homochirality\par}
\vspace{1em}

S. Furkan Ozturk$^{1}$, 
Roger R. Fu$^{2}$,
José A. P. M. Devienne$^{3}$,
Sammy Piascik$^{2}$, 
Perri Thaler$^{2}$,
Aaron T. Kuan$^{3}$,
Richard J. Harrison$^{4}$,
Dimitar D. Sasselov$^{5}$

\vspace{0.75em}

\footnotesize
$^{1}$Division of Geological and Planetary Sciences, California Institute of Technology, Pasadena, CA 91125, USA\\
$^{2}$Department of Earth and Planetary Sciences, Harvard University, Cambridge, MA 02138, USA\\
$^{3}$Yale School of Medicine, Yale University, New Haven, CT 06510, USA\\
$^{4}$Department of Earth Sciences, University of Cambridge, Cambridge CB2 3EQ, UK\\
$^{5}$Department of Astronomy, Harvard University, Cambridge, MA 02138, USA

\end{center}

\setcounter{page}{1} 
\setcounter{section}{0}
\setcounter{figure}{0}
\setcounter{table}{0}
\setcounter{equation}{0}

\renewcommand{\thefigure}{S\arabic{figure}}
\renewcommand{\thetable}{S\arabic{table}}
\renewcommand{\theequation}{S\arabic{equation}}

\newpage

\section*{QDM analysis of the second field of view (QDM2)}

\begin{table}[H]
\centering
\begin{tabular}{lrrr}
\toprule
Quantity & Raw & Dipolarity filter & Dipolarity + $|m|$ filter \\
         &     & ($R^{2} \geq 0.3$, $\mathrm{SNR} \geq 1$) & (plus $|m| \geq 5 \times 10^{-15}$~A\,m$^2$) \\
\midrule
Number of dipoles, $N$                   & 212                   & 171                   & 142                   \\
Inclination ($^\circ$)                   & $+23.3$               & $+26.3$               & $+26.3$               \\
Declination ($^\circ$)                   & $338.4$               & $339.8$               & $339.7$               \\
Net moment amplitude (A\,m$^2$)          & $1.46 \times 10^{-11}$ & $1.49 \times 10^{-11}$ & $1.49 \times 10^{-11}$ \\
\midrule
Watson resultant, $R$                    & $23.10$               & $24.34$               & $24.42$               \\
Watson $R_{\text{crit}}$ ($5\,\%$)       & $23.50$               & $21.11$               & $19.23$               \\
Watson $R_{\text{crit}}$ ($1\,\%$)       & $28.32$               & $25.43$               & $23.17$               \\
Watson test outcome                      & not rejected          & rejected at $5\,\%$   & rejected at $1\,\%$   \\
\midrule
$\%\mathrm{DRM_{moment}}$                & $63.9\,\%$            & $69.9\,\%$            & $70.3\,\%$            \\
$\%\mathrm{DRM}_{\hat n}$        & $10.9\,\%$            & $14.2\,\%$            & $17.2\,\%$            \\
Fisher $\alpha_{95}$ (unit vectors; $^\circ$) & $8.6$             & $9.2$                 & $9.8$                 \\
\bottomrule
\end{tabular}
\caption{QDM2 directional-analysis results under three successive filtering stages. Columns correspond to (i) all detected dipoles, (ii) dipoles passing a dipolarity-based quality filter ($R^{2} \geq 0.3$, $\mathrm{SNR} \geq 1$), and (iii) the additional restriction $|m_i| \geq 5 \times 10^{-15}$~A\,m$^2$. The net vector direction is stable under filtering and agrees, within Fisher $\alpha_{95}$, with both the 2G bulk DRM direction and the QDM1 net sum, reinforcing that the DRM direction is set by the largest dipoles in each population. The unusually high $\%\mathrm{DRM_{moment}}$ ($\sim 70\,\%$) and its factor-${\sim}4$ offset from $\%\mathrm{DRM}_{\hat n}$ ($\sim 17\,\%$) are driven by a single $\sim 10^{-11}$~A\,m$^2$ dipole in this field of view that dominates the moment-weighted sum while leaving the unit-vector statistics essentially unchanged. The Watson randomness test~\cite{Watson1956} compares $R$ to $R_{\text{crit}}$ at the $5\,\%$ and $1\,\%$ levels; the null hypothesis of isotropic directions is marginal in the raw population and is rejected at the $1\,\%$ level after both filters. \%DRM metrics follow \autoref{eq:DRM_definitions}, and $\alpha_{95}$ is computed on unit vectors~\cite{fisher1953}.}
\label{tab:qdm2_filter_table}
\end{table}

\newpage

\begin{figure}[H]
    \centering
    \includegraphics[width=0.95\linewidth]{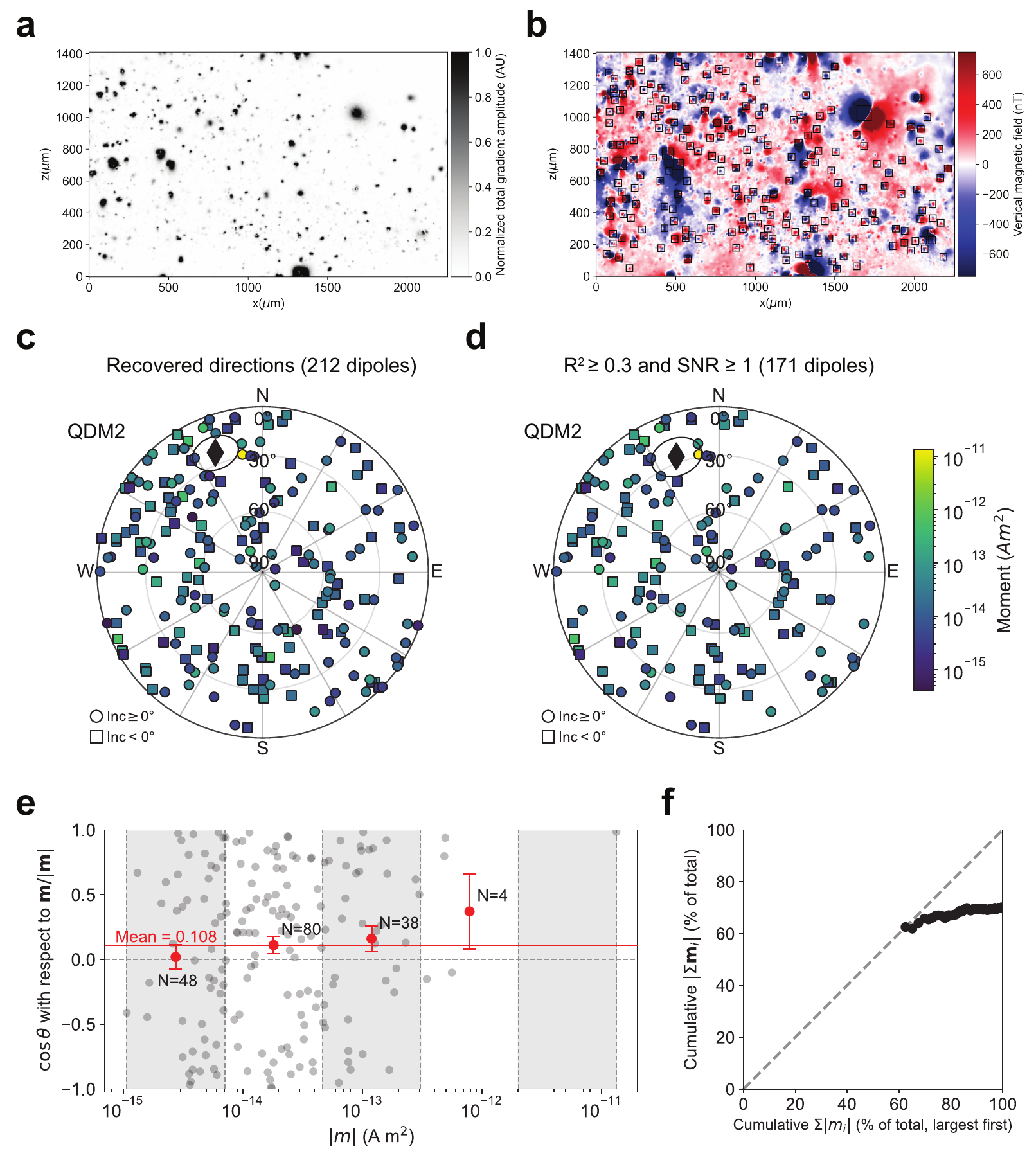}
    \caption{Particle-level QDM analysis of the second field of view (QDM2). 
    \textbf{(a)} Normalized total magnetic gradient amplitude. 
    \textbf{(b)} Vertical magnetic field map upward-continued by $5\,\mu$m; black squares mark the bounding boxes of dipoles recovered by Euler deconvolution.
    \textbf{(c)} Equal-area projection of all $N = 212$ recovered dipole directions, colored by inverted moment magnitude. The black diamond marks the net vector sum ($\mathrm{DRM_{moment}}$), with its Fischer $\alpha_{95}$. 
    \textbf{(d)} Same as (c) after applying the dipolarity quality filter ($R^{2} \geq 0.3$, $\mathrm{SNR} \geq 1$; $N = 171$); the net direction is essentially unchanged. 
    \textbf{(e)} Per-dipole projection $\cos\theta$ onto the net-moment direction as a function of $|m|$. Grey points are individual dipoles; red markers are logarithmic-bin means with standard errors; the red line is the population mean. The largest-$|m|$ bin ($N=4$) shows markedly higher alignment than the smaller-moment bins, consistent with the DRM being set by the largest grains.
    \textbf{(f)} Cumulative fraction of $|\Sigma \mathbf{m}_i|$ as dipoles are added largest-first, plotted against the cumulative fraction of $\Sigma|m_i|$. The curve rises steeply, reaching $\sim 70\%$ of the net vector sum after only a few of the largest dipoles are included.}
    \label{fig:figure_S1}
\end{figure}

\newpage

\section*{Electron microscopy and chemical mapping}

\begin{figure}
    \centering
    \includegraphics[width=\linewidth]{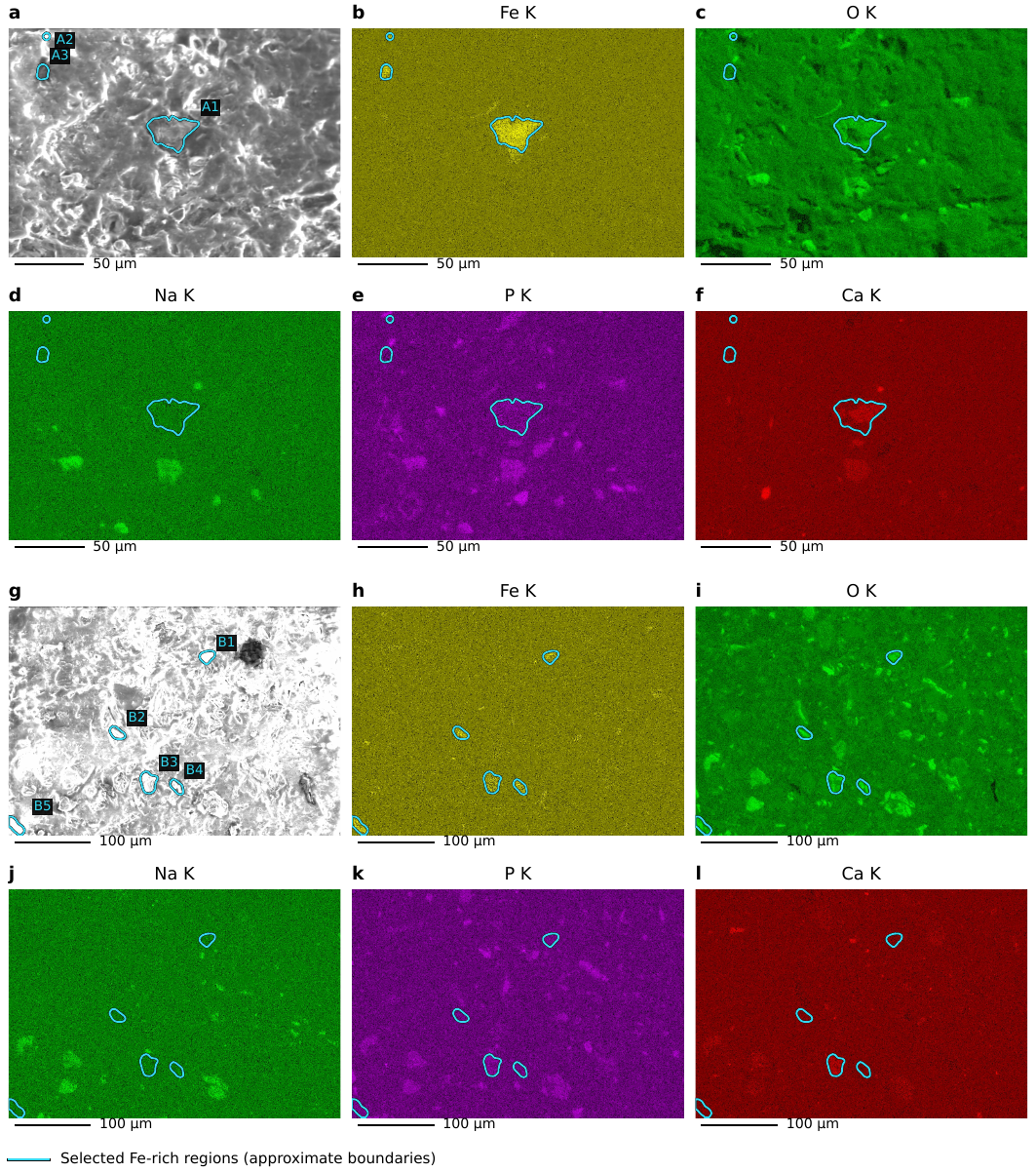}
    \caption{SEM images and EDS elemental maps of lake sediment. Two fields of view are shown in panels (a–f) and (g–l). SEM images (a, g) are accompanied by elemental maps of Fe (b, h), O (c, i), Na (d, j), P (e, k) and Ca (f, l). Cyan outlines mark selected Fe-rich regions, labelled A1–A3 and B1–B5 in the SEM images, with identical boundaries reproduced across the corresponding elemental maps. Several regions show spatially associated Fe and O enrichment, highlighting potential magnetite-bearing material, although these maps alone do not establish mineral identity. Scale bars represent 50 $\mu$m in panels (a–f) and 100 $\mu$m in panels (g–l).}
    \label{fig:figure_S2}
\end{figure}

SEM imaging and EDS characterization were complicated due to preparation of the sediment samples, which were embedded in electrically insulating epoxy and examined without carbon coating. Limited dissipation of the incident electron charge resulted in substantial charging artifacts, producing strong brightness variations that obscured grain morphology (\ref{fig:figure_S2}a and g). These effects made it difficult to distinguish individual particles from the surrounding matrix and to delineate well-defined regions for comparison with the elemental maps. Identifying discrete Fe-rich regions was particularly challenging because much of the Fe map showed diffuse background signal, with relatively few clearly distinguishable areas of local enrichment. Regions with locally enhanced Fe signal were selected and compared with the corresponding O maps. Spatial overlap between Fe enrichment and elevated O signal supported their interpretation as potential magnetite-bearing regions (Figure \ref{fig:figure_S2}b-f and h-l). The correspondence was, nevertheless, not universal: many O-rich areas lacked a clearly distinguishable Fe-rich counterpart. Oxygen enrichment alone was therefore insufficient to identify candidate magnetite-bearing material. Other chemically distinct components were also observed in the sediment, as reflected by the additional Na, P and Ca signals. 

\section*{Hysteresis measurements}

\begin{figure}[H]
    \centering
    \includegraphics[width=\linewidth]{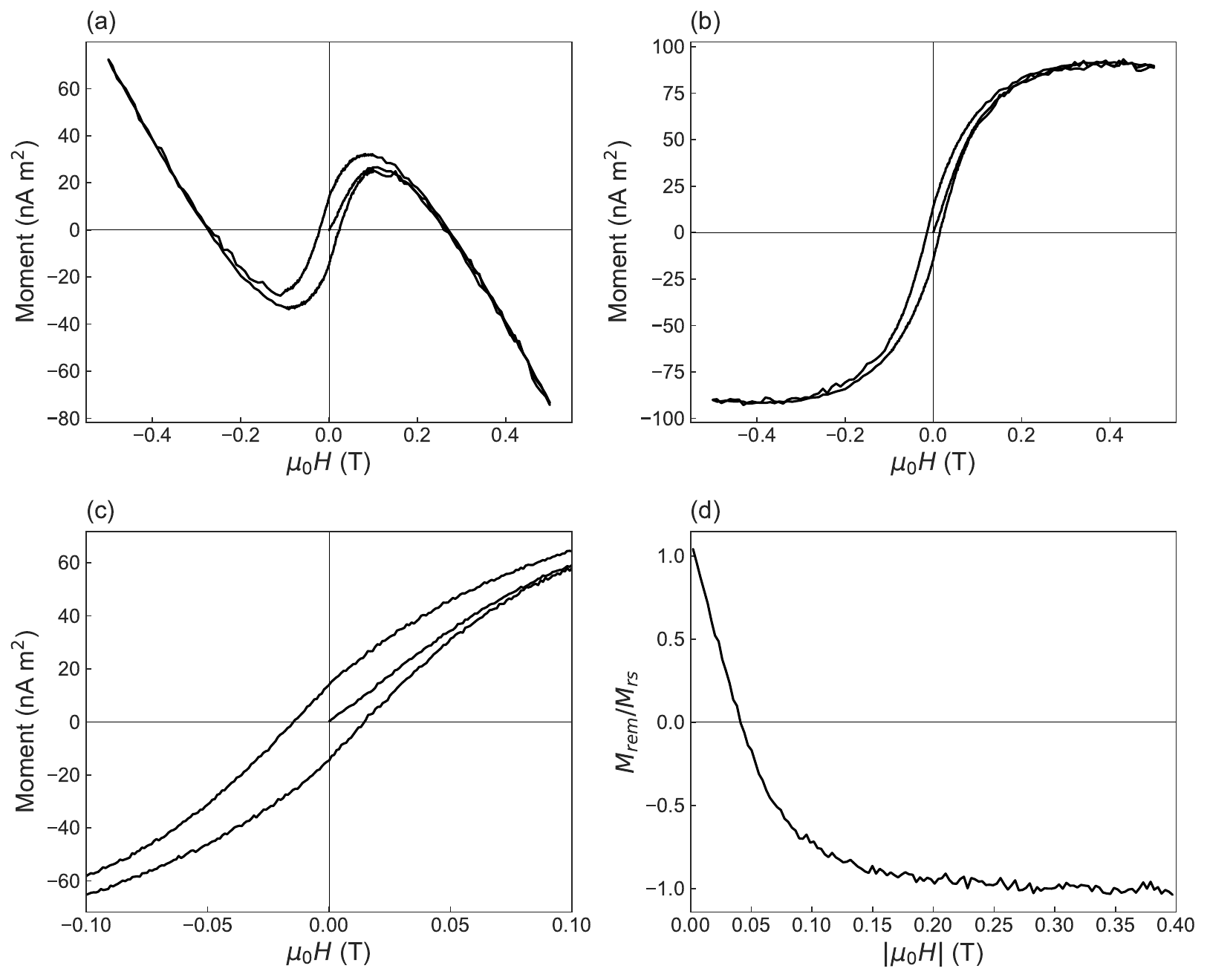}
    \caption{Hysteresis measurements from the lake sediment. (a) Raw hysteresis loop including the initial magnetization curve (only the $\pm$0.5~T range is shown). (b) The same loop as in (a) after the slope correction to remove the strong diamagnetic contribution from the epoxy where the sediment is embedded. (c) Zoomed-in region of the loop shown in (b) within $\pm$0.1~T. (d) DCD curve used to estimate the coercivity of remanence.}
    \label{fig:figure_S3}
\end{figure}

\begin{figure}[H]
    \centering
    \includegraphics[width=0.6\linewidth]{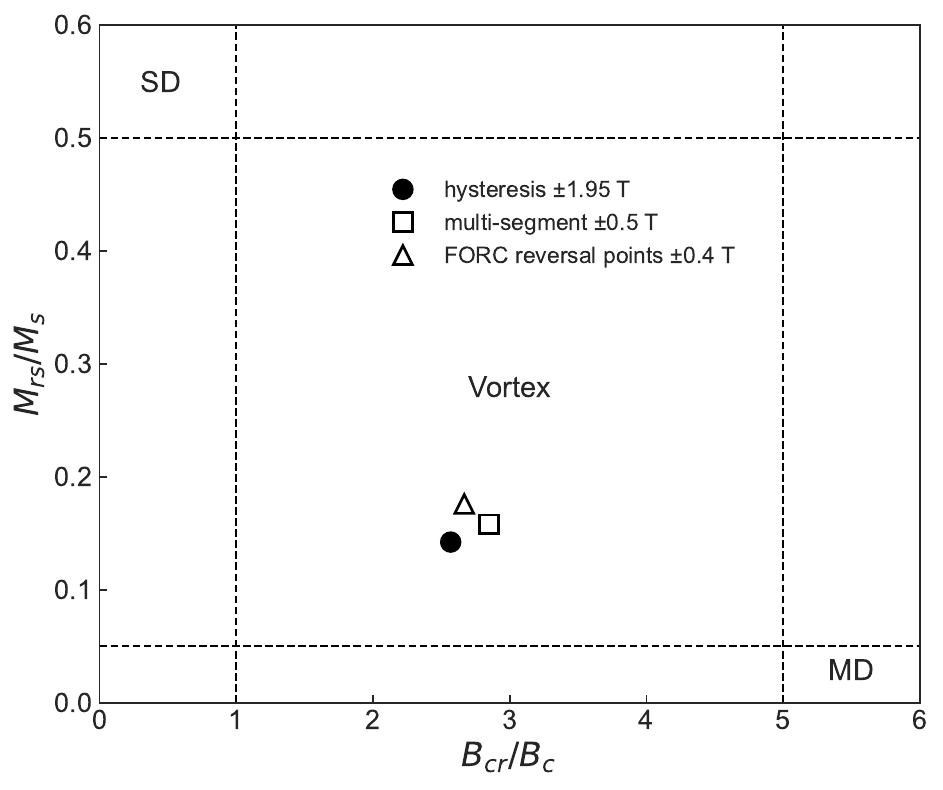}
    \caption{Day plot \cite{Day1977Hysteresis} showing the relation $M_{rs}/M_s$ versus $B_{cr}/B_c$ based on three estimates of the magnetic parameters of the sediment: from the raw full loop ($\pm$1.95~T; black circle), from the multi-segment ($\pm$0.5~T; square) and from the FORC distribution ($\pm$0.4~T; triangle). The three estimates are closely clustered within the vortex size range, indicating the predominance of this domain structure in the bulk magnetic behavior. }
    \label{fig:figure_S4}
\end{figure}

\begin{figure}[H]
    \centering
    \includegraphics[width=\linewidth]{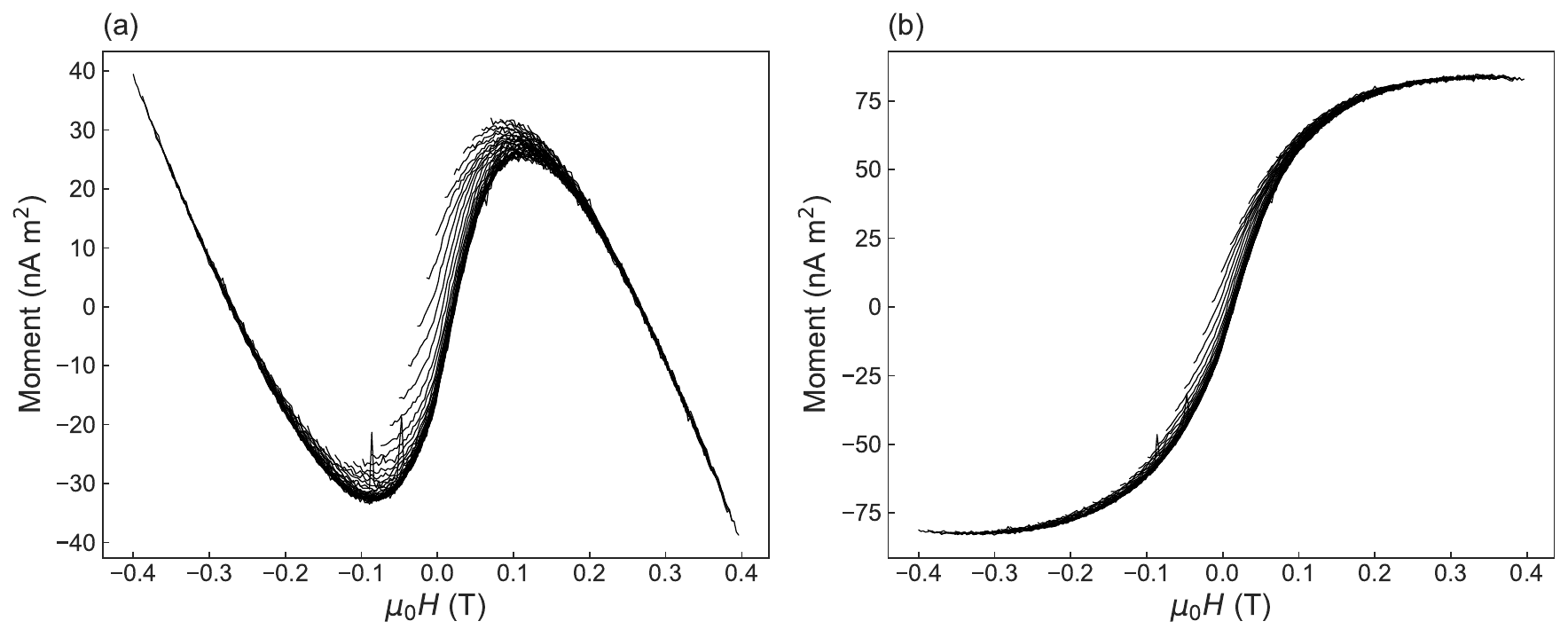}
    \caption{First Order Reversal Curves (FORC). (a) Raw partial loops with strong diamagnetic contribution from the epoxy. (b) Slope-corrected loops after diamagnetic contribution removal. Slope-corrected loops were used for processing using the FORCinel software \cite{Harrison2008}.}
    \label{fig:figure_S5}
\end{figure}

Additional hysteresis measurements are shown in Figures \ref{fig:figure_S3} and \ref{fig:figure_S4}. These include raw hysteresis loops before and after the slope correction to remove the strong diamagnetic contribution from the epoxy where the sediment is embedded. From these measurements, the magnetic parameters obtained for the sediments were saturation magnetization $M_s=1.27\times10^{-7}\mathrm{Am^2}$, saturation remanence magnetization $M_{rs}=1.43\times10^{-8}\mathrm{Am^2}$ and coercivity $B_c=14.6$~mT. From direct current demagnetization (DCD) curves the remanence coercivity $B_{cr}=41.6$~mT was estimated. With these values, the relation $M_{rs}/M_s$ versus $B_{cr}/B_c$ was used to estimate the sediment overall magnetic domain behavior using the so-called Day diagram \cite{Day1977Hysteresis, Dunlop2002DayPlot}. As shown in Figure \ref{fig:figure_S4}, the three estimates of the magnetic parameters of the sediment (i.e., from the $\pm$1.95~T full loop, from the multi-segment within $\pm0.5$~T and from the FORC distribution) are closely clustered within the vortex range, in agreement with the processed FORC (Figure \ref{fig:fig4} and Figure \ref{fig:figure_S5}) that indicates vortex structures (associated with isolated grains and/or intact MTB chains) to dominate the magnetic signal in the sediment.

\newpage

\newpage

\newpage

\section*{Kinetic model of DRM acquisition}

We treat each QDM dipole as a rigid spherical carrier of hydrodynamic volume $V_h$ carrying a fixed magnetic moment $\mathbf{m}$ (magnitude $m$). In our samples the carriers are not free magnetite crystals: the SD magnetite occurs as inclusions within fine sediment, so each carrier is a composite of magnetite and non-magnetic sediment, and it sits in a matrix of fine sediment and resin rather than in pure resin. We assume that the carriers do not interact and that their orientations are random right after dispersion. We first derive how a single carrier aligns, and then evaluate the relevant timescales for our resin.

\subsection*{Kinetic alignment of a single carrier}

Let $\theta$ be the angle between $\mathbf{m}$ and the applied field $\mathbf{B}$ (magnitude $B$), and let the surrounding liquid have viscosity $\eta$. The carrier rotates under three torques: the magnetic torque, which turns $\mathbf{m}$ toward $\mathbf{B}$; viscous drag, which opposes rotation; and random thermal kicks from the liquid. Newton's law for rotation gives the rotational Langevin equation
\begin{equation}
J\,\frac{\mathrm{d}^2\theta}{\mathrm{d}t^2} = -mB\sin\theta - \zeta\,\frac{\mathrm{d}\theta}{\mathrm{d}t} + \xi(t),
\qquad \zeta = 6\eta V_h ,
\label{eq:S-full}
\end{equation}
where $J$ is the carrier's moment of inertia, $-mB\sin\theta$ is the magnetic torque (the derivative of the magnetic energy $U = -mB\cos\theta$), $\zeta$ is the rotational drag coefficient of a sphere, and $\xi(t)$ is a random torque with zero mean. Drag and thermal kicks both arise from collisions with the same liquid, so their strengths are linked: by the fluctuation--dissipation theorem, the mean square of $\xi$ is proportional to $k_\mathrm{B}T\,\zeta$, where $k_\mathrm{B}$ is the Boltzmann constant and $T$ the temperature.

Inertia relaxes in the time $J/\zeta = \rho r^2/15\eta$, where $\rho$ and $r$ are the carrier's density and radius. For $\rho\approx2600$~kg\,m$^{-3}$, $r = 12.5~\mu$m and $\eta = 0.15$~Pa\,s this is below a microsecond, far shorter than any other timescale here. The carrier therefore turns at the rate the torques allow without accelerating, and \eqref{eq:S-full} reduces to the rotational Langevin equation in the overdamped limit

\begin{equation}
\zeta\,\frac{\mathrm{d}\theta}{\mathrm{d}t} = -mB\sin\theta + \xi(t).
\label{eq:S-overdamped}
\end{equation}

Because $\zeta$ multiplies the left-hand side and also sets the strength of $\xi$, dividing \eqref{eq:S-overdamped} by $\zeta$ shows that viscosity acts only as a clock: the carriers pass through the same sequence of states whatever $\eta$ is, faster in a thin liquid and slower in a thick one. For a viscosity $\eta(t)$ that rises as the resin cures and hardens at time $t_h$, the elapsed clock time is $\int\mathrm{d}t/\zeta = I/6V_h$, with the viscous time integral
\begin{equation}
I = \int_0^{t_h}\frac{\mathrm{d}t}{\eta(t)} .
\label{eq:S-I}
\end{equation}
The outcome then depends on two numbers. The first,
\begin{equation}
\alpha = \frac{mB}{k_\mathrm{B}T} = \frac{m}{m_T},\qquad m_T = \frac{k_\mathrm{B}T}{B},
\end{equation}
compares magnetic and thermal energy; $m_T\approx8\times10^{-17}$~A\,m$^2$ at $B = 58.8~\mu$T and $T = 338$~K is the moment whose magnetic energy equals the thermal energy. The second,
\begin{equation}
\Phi = \frac{mB\,I}{6V_h},
\label{eq:S-Phi}
\end{equation}
measures how far the field can turn the carrier before rotation stops. At constant viscosity it is simply $\Phi = t/\tau_B$, where $\tau_B = 6\eta V_h/mB$ is the time the magnetic torque needs to turn the carrier through about one radian against drag. Their ratio, $\Phi/\alpha = k_\mathrm{B}T\,I/6V_h = \int D_r\,\mathrm{d}t$, is the rotational diffusion accumulated during the cure, with $D_r = k_\mathrm{B}T/\zeta$ the rotational diffusion coefficient; thermal rotation can randomize the moments only if this reaches order one.

If the carriers rotate long enough ($\Phi\gtrsim1$ and $\Phi\gtrsim\tfrac12\ln\alpha$, a few alignment times), they reach thermal equilibrium, the Boltzmann distribution $p\propto e^{-U/k_\mathrm{B}T}$, whose mean alignment is the Langevin function $\langle\cos\theta\rangle = \mathcal{L}(\alpha) = \coth\alpha - 1/\alpha$. This depends only on $\alpha$: viscosity sets how fast equilibrium is approached but not where it lies, so it drops out of the equilibrium result. If rotation is arrested earlier, the frozen state depends on how far the approach has progressed, that is, on $\Phi$, and therefore on viscosity through $I$. Because the fluctuation--dissipation theorem holds at every instant, the entire viscosity history enters only through $I$.

\subsubsection*{Analytical solution for negligible thermal noise}
 
When $\Phi/\alpha\ll1$, as for all our carriers (see below), the thermal torque can be neglected and \eqref{eq:S-overdamped} becomes $6\eta(t)V_h\,\mathrm{d}\theta/\mathrm{d}t = -mB\sin\theta$. Dividing by $6\eta V_h\sin\theta$, using $\mathrm{d}\theta/\sin\theta = \mathrm{d}\ln\tan(\theta/2)$ and integrating from the initial angle $\theta_0$ gives
\begin{equation}
\tan\frac{\theta}{2} = \tan\frac{\theta_0}{2}\,e^{-\Phi},\qquad
\Phi = \frac{m}{m^*},\qquad
m^* = \frac{6V_h}{BI}.
\label{eq:S-traj}
\end{equation}
The characteristic moment $m^*$ is the moment for which $\Phi = 1$: carriers with $m\ll m^*$ barely turn before they are locked in, and carriers with $m\gg m^*$ end up fully aligned. Written in terms of $x = \cos\theta$, using $\tan^2(\theta/2) = (1-x)/(1+x)$, \eqref{eq:S-traj} becomes
\begin{equation}
x = \frac{x_0 + c}{1 + c\,x_0},\qquad c = \tanh\Phi .
\end{equation}

Random initial orientations mean that $x_0$ is uniformly distributed on $[-1,1]$. Changing variables from $x_0$ to $x$ gives the frozen distribution
\begin{equation}
p(x) = \frac{1}{2\,(\cosh\Phi - x\sinh\Phi)^{2}},
\end{equation}
the spherical Cauchy distribution. It has a heavier tail toward $x = -1$ than the Fisher distribution of equilibrium, because carriers that start nearly antiparallel to the field feel almost no torque. Averaging $x$ over $x_0$, after writing $(x_0+c)/(1+cx_0) = 1/c - (1-c^2)/[c\,(1+cx_0)]$, gives
\begin{equation}
\langle\cos\theta\rangle = \frac1c - \frac{1-c^2}{2c^2}\ln\frac{1+c}{1-c},
\end{equation}
which, with $c = \tanh\Phi$, simplifies to the kinetic Langevin function
\begin{equation}
\boxed{\;\langle\cos\theta\rangle = \mathcal{K}(\Phi) = \coth\Phi - \frac{\Phi}{\sinh^2\Phi}\;}
\label{eq:S-K}
\end{equation}
with limits
\begin{equation}
\mathcal{K}(\Phi) \approx \tfrac23\Phi - \tfrac{4}{45}\Phi^3\quad(\Phi\ll1),
\qquad
\mathcal{K}(\Phi) \approx 1 - (4\Phi-2)\,e^{-2\Phi}\quad(\Phi\gg1).
\label{eq:S-limits}
\end{equation}
The rest of the model only requires evaluating $\mathcal{K}(m_i/m^*)$ for each dipole $i$.

\subsubsection*{Comparison with the equilibrium model} 

For weak alignment, $\mathcal{L}(\alpha)\approx\alpha/3$ and $\mathcal{K}(\Phi)\approx2\Phi/3$, so the kinetic result equals an equilibrium one with $\alpha$ replaced by $2\Phi$. Equivalently, the frozen disorder acts like an effective temperature $k_\mathrm{B}T_\mathrm{eff} = m^*B/2 = 3V_h/I$, set by carrier size and viscosity history rather than by $T$. The two functions are compared in \autoref{fig:fig3}c. Restoring the thermal torque and solving \eqref{eq:S-overdamped} to leading order in it gives
\begin{equation}
\langle\cos\theta\rangle \approx \mathcal{K}(\Phi)\left(1 - \frac{\Phi}{\alpha}\right),
\qquad \frac{\Phi}{\alpha} = \frac{m_T}{m^*},
\end{equation}
which agrees with a numerical solution of the full equation. For the fitted $m^*$, $\Phi/\alpha = 1.9\times10^{-4}$, so the correction is negligible.

\subsection*{Rotation timescales in epoxy}

EPO-TEK 301 epoxy has a viscosity of $\eta_0 = 0.1$--0.2~Pa\,s at 23\,$^\circ$C. The samples were cured at 65\,$^\circ$C ($T = 338$~K), where the resin starts thinner, and hardened within $t_h\approx2$~h. Because $1/\eta$ is largest while the resin is still fluid, $I$ \eqref{eq:S-I} is dominated by the early part of the cure, and the later rise in viscosity adds little. For a conservative estimate we take $\eta_0 = 0.15$~Pa\,s, the 23\,$^\circ$C value, and let the viscosity rise exponentially by a factor of $10^{4}$ over the 2~h, $\eta(t) = \eta_0\,e^{t/t_c}$ with $t_c = t_h/\ln10^{4}\approx13$~min. This gives
\begin{equation}
I \approx \frac{t_c}{\eta_0}\approx5\times10^{3}~\mathrm{Pa^{-1}}.
\end{equation}
With the lower starting viscosity expected at 65\,$^\circ$C (e.g.\ 0.03~Pa\,s), $I$ would be about five times larger.

\autoref{tab:S-timescales} gives the accumulated thermal rotation $\Phi/\alpha$ and the alignment parameter $\Phi$ for a carrier with $m = 10^{-14}$~A\,m$^2$, near the middle of our distribution, rotating freely in the neat resin. It also lists the time the carrier needs to align almost completely, $3\tau_B$ at the starting viscosity, after which $\langle\cos\theta\rangle = \mathcal{K}(3) = 0.97$. A carrier that completes this within the $\sim13$~min before the resin stiffens reaches near-saturated alignment; one that cannot is frozen part way.

\begin{table}[H]
\centering
\begin{tabular}{rcccc}
\toprule
Carrier diameter & $\int D_r\,\mathrm{d}t$ & $\Phi$ & $\mathcal{K}(\Phi)$ & Time to align ($3\tau_B$) \\
\midrule
1~$\mu$m  & 8    & $1.0\times10^{3}$ & 1.00 & 2~s \\
5~$\mu$m  & 0.06 & 7.8 & 1.00 & 5~min \\
10~$\mu$m & 0.008 & 1.0 & 0.58 & 40~min \\
25~$\mu$m & 0.0005 & 0.06 & 0.04 & 10~h \\
50~$\mu$m & 0.00006 & 0.008 & 0.01 & 3.5~days \\
\bottomrule
\end{tabular}
\caption{Rotation of a carrier with $m = 10^{-14}$~A\,m$^2$ rotating freely in neat EPO-TEK 301 during the cure ($I = 5\times10^{3}$~Pa$^{-1}$, $\eta_0 = 0.15$~Pa\,s for the alignment time, $B = 58.8~\mu$T, $T = 338$~K).}
\label{tab:S-timescales}
\end{table}

Thermal rotation is small for every carrier of a few micrometres or more, so thermal randomization cannot explain the weak alignment; a 1~$\mu$m carrier would equilibrate thermally, but with $\alpha\approx130$ it would then be fully aligned. Free rotation in the neat resin would fully align carriers up to about 5~$\mu$m within minutes, partly align 10~$\mu$m carriers, and barely turn 25--50~$\mu$m carriers. We return to this when comparing the model with the data.

\subsection*{Population metrics}

The two DRM metrics measure how well the dipoles add up along the field. For weights $w_i$,
\begin{equation}
\%\mathrm{DRM} = \frac{|\mathbf{S}|}{\sum_i w_i},\qquad \mathbf{S} = \sum_i w_i\hat{\mathbf m}_i ,
\end{equation}
with $w_i = m_i$ for $\%\mathrm{DRM_{moment}}$ and $w_i = 1$ for $\%\mathrm{DRM}_{\hat n}$. Each dipole contributes on average $w_i\mathcal{K}_i$ along $\mathbf{B}$, where $\mathcal{K}_i = \mathcal{K}(m_i/m^*)$, so for a very large population the metrics reduce to weighted averages of $\mathcal{K}$, exactly as in the equilibrium model with $\mathcal{L}$ replaced by $\mathcal{K}$:
\begin{equation}
\%\mathrm{DRM_{moment}}\to\frac{\sum_i m_i\,\mathcal{K}_i}{\sum_i m_i},\qquad
\%\mathrm{DRM}_{\hat n}\to\frac1N\sum_i\mathcal{K}_i .
\end{equation}
For our $N = 191$ QDM1 dipoles, the random scatter of individual orientations also adds to $|\mathbf{S}|$: as in a random walk, $N$ randomly oriented unit vectors sum to a length of about $\sqrt N$, not zero. Because the orientations are independent and $|\hat{\mathbf m}_i| = 1$, the mean square length is exactly
\begin{equation}
\mathbb{E}\,|\mathbf{S}|^2 = \Bigl(\sum_i w_i\,\mathcal{K}_i\Bigr)^2 + \sum_i w_i^2\left(1-\mathcal{K}_i^2\right),
\label{eq:S-rms}
\end{equation}
where the first term is the aligned signal and the second the random-walk contribution. We take $\sqrt{\mathbb{E}|\mathbf S|^2}/\sum_i w_i$ as the predicted metric. For completely random orientations ($\mathcal{K}_i = 0$) this gives a baseline of $\sqrt{\sum_iw_i^2}/\sum_iw_i$: 7\,\% ($=1/\sqrt N$) for $\%\mathrm{DRM}_{\hat n}$ and 14\,\% for $\%\mathrm{DRM_{moment}}$, which is higher because a few large dipoles dominate the sum.

\subsection*{Fit to the data and model interpretation}

The only free parameter, $m^*$, is fitted to the five per-bin mean values of $\cos\theta$, with each bin represented by its mean moment. The fit gives $m^* = 4.3\times10^{-13}$~A\,m$^2$ (68\,\% range $3.2$--$6.2\times10^{-13}$~A\,m$^2$) and $\chi^2 = 2.8$ for 4 degrees of freedom (\autoref{fig:fig3}d). A single $m^*$ describes all bins well: a more flexible dependence on moment barely improves the fit, and inverting $\mathcal{K}$ for each bin separately gives carrier sizes that agree within their uncertainties (\autoref{tab:S-Vh}). 

Using this $m^*$, the model predicts $\%\mathrm{DRM_{moment}} = 26\,\%$ and $\%\mathrm{DRM}_{\hat n} = 10\,\%$, close to the measured 24\,\% and 10\,\%. The measured alignment of the largest dipoles may be slightly overestimated, because each contributes to the net DRM direction against which it is measured, but this effect is within the error bars.

\begin{table}[H]
\centering
\begin{tabular}{rcccc}
\toprule
$\langle|m|\rangle$ (A\,m$^2$) & Measured $\langle\cos\theta\rangle$ & Required $\Phi$ & $V_h$ (m$^3$) & Equivalent diameter ($\mu$m) \\
\midrule
$3.7\times10^{-15}$ & $-0.10\pm0.10$ & -- & -- & $>13$\textsuperscript{a} \\
$1.2\times10^{-14}$ & $\phantom{-}0.08\pm0.07$ & 0.12 & $5\times10^{-15}$ & 21 (17--43) \\
$3.8\times10^{-14}$ & $\phantom{-}0.09\pm0.07$ & 0.14 & $1.4\times10^{-14}$ & 30 (25--50) \\
$1.2\times10^{-13}$ & $\phantom{-}0.23\pm0.09$ & 0.35 & $1.7\times10^{-14}$ & 32 (29--38) \\
$3.9\times10^{-13}$ & $\phantom{-}0.42\pm0.19$ & 0.67 & $3.0\times10^{-14}$ & 39 (33--48) \\
\midrule
\multicolumn{3}{l}{Single fit, $m^* = 4.3\times10^{-13}$~A\,m$^2$} & $2.2\times10^{-14}$ & 35 \\
\bottomrule
\end{tabular}
\caption{Carrier size implied by the measured alignment in each moment bin. $\Phi$ is obtained by inverting $\mathcal{K}$, and $V_h = mBI/6\Phi$ assumes purely viscous arrest with a common $I = 5\times10^{3}$~Pa$^{-1}$. Diameters are for the equivalent sphere, with $\pm1\sigma$ ranges in parentheses. \textsuperscript{a}The alignment of the smallest-moment bin is consistent with zero, so $\Phi$ is not constrained; the value given is a lower bound from the $2\sigma$ upper limit of the alignment.}
\label{tab:S-Vh}
\end{table}

\paragraph{Carrier composition.} A 20--100~nm SD magnetite crystal carries only $2\times10^{-18}$--$2.5\times10^{-16}$~A\,m$^2$, so no single crystal can supply the measured moments, and each carrier must contain many. Such crystals do not settle in isolation but attach to clay and fine silt~\cite{TauxeSteindorf2006, KatariBloxham2001}, consistent with our images of magnetite inclusions within fine sediment. The carrier's size, and hence its drag, is then set by the host sediment, and its moment by the magnetite it holds. Writing $m = f\,\phi\,M_sV_h$, with $M_s = 480$~kA\,m$^{-1}$ the saturation magnetization of magnetite, $\phi$ the volume fraction of magnetite in the carrier and $f\le1$ the degree to which the crystal moments point the same way, the largest measured moment ($\sim4\times10^{-13}$~A\,m$^2$) requires $8\times10^{-19}/f$~m$^3$ of magnetite. For $f = 1$ this is about $10^{4}$ crystals of 50~nm, equivalent to a single 1.2~$\mu$m magnetite sphere; for randomly oriented crystals the net moment grows only as the square root of their number, and about $10^{8}$ would be needed. The largest carriers must therefore hold their crystals in coherent clusters or chains, consistent with intact magnetosome chains. This magnetite is a tiny fraction of the carrier volume, $f\phi\approx2\times10^{-3}$, $1\times10^{-4}$ and $1\times10^{-5}$ for 10, 25 and 50~$\mu$m carriers, all below the 63~$\mu$m sieve size that bounds every grain in the sample.

\paragraph{What $m^*$ implies about rotation.} If $m^*$ reflects viscous arrest alone, \eqref{eq:S-traj} gives the viscous time integral the carriers experienced as $I_\mathrm{eff} = 6V_h/m^*B$. For 50~$\mu$m carriers, $I_\mathrm{eff}\approx1.6\times10^{4}$~Pa$^{-1}$, which the resin itself could supply if its starting viscosity at 65\,$^\circ$C is about 0.05~Pa\,s. For 25 and 10~$\mu$m carriers, $I_\mathrm{eff}\approx2\times10^{3}$ and $1.2\times10^{2}$~Pa$^{-1}$, about 3 and 40 times below our neat-resin estimate. Curing cannot close this gap, because $I$ is set by the early, fluid stage, but a dense sediment--resin mixture is more viscous than the neat resin, and grain contacts and shape-controlled torques can further restrict rotation.

\paragraph{Limitations.} The model treats each carrier as a sphere with drag coefficient $6\eta V_h$. Real carriers are irregular or elongated; for the same volume, an elongated carrier has a larger and axis-dependent rotational drag, which changes the $V_h$ inferred from $m^*$ but not the form of $\mathcal{K}(\Phi)$. Non-spherical shape also gives rise to hydrodynamic and gravitational torques during settling, which may exceed the magnetic torque for most of our carriers (discussed in the main text). Because these torques likewise produce an alignment that grows in proportion to the moment, their effect cannot be separated from viscous arrest by our data and is absorbed into the fitted $m^*$, which is best read as the moment at which the magnetic torque overcomes all non-magnetic constraints on rotation. The model also assumes a single $m^*$ for all carriers, no interactions between carriers, and random orientations when the field starts to act.

\end{document}